\documentclass[reprint,aps,prb,english,superscriptaddress,nofootinbib]{revtex4-2}

\usepackage{graphicx}
\usepackage{physics}
\usepackage{bm,amssymb,amsmath,mathrsfs,amstext,latexsym}

\usepackage[usenames,dvipsnames]{xcolor}
\usepackage[colorlinks=true,citecolor=blue,urlcolor=blue,linkcolor=blue]{hyperref}
\usepackage[normalem]{ulem}

\newcommand{\rd}{\mathrm{d}}
\newcommand{\rhodec}{\rho_{\mathrm{dec}}}
\newcommand{\rhores}{\rho_{\mathrm{res}}}
\newcommand{\HLevy}{H_{\text{LRP}}}
\newcommand{\HRRG}{H_{\text{RRG}}}
\newcommand{\HXXZ}{H_{\text{XXZ}}}

\begin{document}

\title{Resonance Proliferation Across Localization Transitions}

\author{Carlo Vanoni}
\email{cvanoni@princeton.edu}
\altaffiliation{equal contribution}
\affiliation{Department of Physics, Princeton University, Princeton, New Jersey, 08544, USA}

\author{David M. Long}
\email{dmlong@stanford.edu}
\altaffiliation{equal contribution}
\affiliation{Department of Physics, Stanford University, Stanford, California 94305, USA}

\author{Anushya Chandran}
\affiliation{Department of Physics, Boston University, Boston, Massachusetts 02215, USA}

\date{\today}

\begin{abstract}
    Models of many-body localization (MBL) exhibit slow numerical drifts towards delocalization with increasing system size, for which no satisfactory theory exists.
    Numerics indicates that these drifts are driven by the proliferation of many-body resonances at intermediate disorder strengths.
    We develop a statistical method to predict the distribution of resonance oscillation frequencies which captures how the formation of resonances at larger frequency scales subsequently affects the formation of resonances at lower frequencies.
    Working within the statistical Jacobi approximation (SJA), we derive a flow equation for a power-law exponent \(\theta(w)\) characterizing the density of resonances at frequency scale \(w\).
    A localized phase is described by a line of fixed points with \(\theta(w)>0\), while an instability of the flow signals resonance proliferation and the onset of thermalization.
    The predicted \(\theta(w)\) matches numerics on the Anderson model on random regular graphs and the L\'evy-Rosenzweig-Porter random matrix ensemble, both of which host resonance-driven delocalization transitions. We further connect the flow to eigenstate properties such as the participation ratio and to dynamical observables such as the return probability.
    The predicted \(\theta(w)\) also matches what is numerically measured in real-space models of MBL at intermediate disorder strengths, representing a significant step towards explaining the finite-size drifts observed in MBL.
\end{abstract}

\maketitle

\section{Introduction}
\label{sec:Intro}

Many-body localization (MBL) purports a robust phase of interacting quantum matter in which generic initial states fail to reach thermal equilibrium under the system's own dynamics~\cite{Anderson1958absence,Gornyi2005,Basko06,oganesyan2007localization,Pal10,Imbrie2016,Schreiber2015,Rubio2019}.
Sufficiently strong random potentials are believed to cause this failure, localizing information about the initial conditions to finite regions of space, so that the system fails to satisfy the foundational ergodic hypothesis of statistical mechanics~\cite{nandkishore2015many,Imbrie17,Abanin2019colloquium,sierant24MBLreview}.

Empirical numerical studies, however, report slow drifts towards delocalization with increasing system size~\cite{Pal10,Luitz15,vsuntajs2020quantum,Sierant20Thouless,Taylor2021,Sels2021Dynamical,Sels2022avalanches,sierant2022challenges, abanin2021distinguishing, panda2020can}.
For instance, in spin chains, spectral probes of ergodicity, such as the \(r\)-statistic~\cite{oganesyan2007localization}, drift towards values characteristic of the ergodic phase at disorder strengths that are several times the bare single-particle bandwidth. Consequently, the disorder strength required for localization increases with system size, empirically in proportion to the size~\cite{vsuntajs2020quantum}.

Similarly, direct measurements of the thermalization time have found an exponential increase with disorder strength, but no clear divergence at a finite disorder strength~\cite{vsuntajs2020quantum,Sierant20Thouless,Sels2021Dynamical}.
These results establish a broad regime of slow finite-size and finite-time dynamics~\cite{Luitz2016Extended,Sierant20Thouless}, which is often interpreted as \emph{prethermal}, where thermalization eventually occurs but only at very late times and very large system sizes~\cite{Crowley2022Constructive,Morningstar2022Avalanches,Long2023Phenomenology}.
The extent of this prethermal regime has generated skepticism regarding the stability of the infinite-volume MBL phase~\cite{vsuntajs2020quantum,Sels2021dynamicalobstruction,sierant2022challenges}.
More immediately, however, we are prompted to understand the physical mechanisms underlying the slow scale dependence of prethermal MBL.

Two potential instabilities to MBL have been identified. At extremely large system sizes, \emph{thermal avalanches}~\cite{DeRoek2017Stability,Luitz2017SmallBath,Crowley:2020ve} can be nucleated by rare weakly-disordered regions.
The weakly disordered patch acts as a thermal bath for surrounding degrees of freedom, which then thermalize and act as a bath for more distant spins, and so on.
This instability is believed to destabilize the infinite-volume MBL phase when the disorder strength is lowered below a (large) threshold value.
However, the slow drifts observed at numerically accessible system sizes are more plausibly attributed to \emph{many-body resonances}~\cite{Gopalakrishnan2015,Khemani2017Critical,villalonga2020eigenstateshybridizelengthscales,Crowley2022Constructive,Garratt2021,Long2023Phenomenology,Ha2023Resonances,jiang2026chargetransportcapacityprobe,Ghosh2022Resonance,Colbois2024Interaction,Laflorencie2025Cat}, which precede the thermal avalanches at intermediate disorder strengths, and which are the main topic of this paper.

Many-body resonances can be understood within a dynamical picture, as follows. Initial product states related by \(n\) spin flips require going to roughly the \(n\)th order of perturbation theory before a non-zero matrix element connecting them can be found. The matrix element is exponentially small in \(n\), but the two initial product states may happen to have an energy separation that is even smaller than the matrix element.
This results in slow Rabi oscillations between these many-body states, which is the direct signature of many-body resonances.
If these resonating many-body states were to find other states to resonate with, which in turn resonate again, the process can run away, and the local pattern of polarization in the initial state will slowly decay~\cite{Long2023Phenomenology}.

\begin{figure*}
    \centering
    \includegraphics[width=\textwidth]{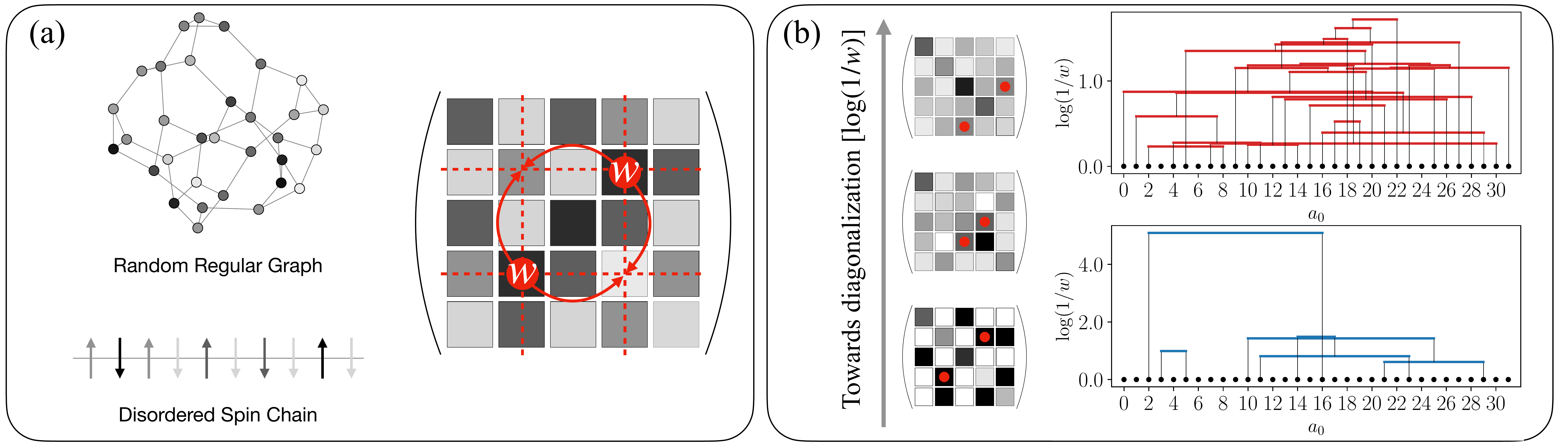}
    \caption{(a)~Schematic representation of two of the models we study using the Jacobi algorithm, namely the Anderson model on random regular graphs (top left) and a disordered quantum spin chain (bottom left). The right panel shows the action of the Jacobi algorithm on a matrix, with the grayscale representing the size of the matrix element. At each iteration, the current largest element, whose value is denoted by $w$ and highlighted in red, is zeroed by a $2$-level rotation. This rotation affects the rows and columns containing the element $w$ (dashed red lines). (b)~As the Jacobi procedure is iterated, the off-diagonal norm and the largest off-diagonal element are reduced, so that the matrix is increasingly diagonal (left). The large Jacobi rotations are identified as \emph{resonances}. Resonances are illustrated by dendrogram arches linking state indices $a_0$, where the height of the arch indicates the value of \(w\) associated with the resonance (right). In the localized phase (lower, blue), there are few resonances, while in the ergodic phase (upper, red), there are many, including resonances formed between states already participating in a resonance. 
    }
    \label{fig:Fig1}
\end{figure*}

Many-body resonances are naturally described in Hilbert space. 
Indeed, they also appear in models of \emph{Hilbert space localization}~\cite{tikhonov2021AndersonMBL}.
Such models are usually defined on expander graphs (to model the local tree-like structure of many-body Hilbert space), with potential disorder on each site and hopping amplitudes between sites. The disorder causes eigenstates of the Hamiltonian to localize on sites. This can be quantified by suitable moments of the eigenstates in the site basis, e.g., the participation ratio (PR). A bounded PR indicates graph/Hilbert space localization, while a growing PR indicates delocalization.

The relationship between MBL and Hilbert space localization is subtle. Even assuming MBL's stability, typical MBL eigenstates are not localized in the basis chosen by the potential disorder~\cite{Deluca13}. Their PR scales with the Hilbert space dimension \(N\) as \(N^c\), where \(0<c<1\). Nevertheless, a toy model related to MBL may show a genuine Hilbert space localization transition. Consider an ideal (possibly fine-tuned) MBL chain and define a graph whose sites represent the eigenstates of the chain. A local perturbation to the chain, say on a single site, induces hopping amplitudes between all the sites. When the localization length of the chain is too long, this perturbation introduces a number of resonances between sites of the Hilbert space graph that diverges with \(N\), and a diverging PR of the perturbed eigenstates. On the other hand, a short enough localization length has been predicted to result in only a finite number of resonances per state~\cite{Crowley2022Constructive}. The stability of the ideal MBL chain to generic local perturbations thus maps to a problem of Hilbert space localization.

How do we track resonance proliferation in a reference Hilbert space? 
Ref.~\cite{Long2023Phenomenology} introduced the \emph{Jacobi algorithm} for this purpose: starting in the disorder-defined computational basis, it iteratively diagonalizes \(H\) via successive two-state rotations that decimate large off-diagonal matrix elements [\autoref{fig:Fig1}(a)], eventually producing the eigenbasis. Operationally, we call a Jacobi step resonant when it produces a large two-level rotation, which occurs when the instantaneous diagonal-energy mismatch is comparable to or smaller than the off-diagonal matrix element coupling the two states. The size of the corresponding decimated matrix element---denoted by \(w\)---gives the corresponding Rabi frequency for this resonance. This identification applies in the dilute-resonance regime, where individual two-state resonances can still be resolved. Once resonances overlap and the matrix enters the dense regime, the scale \(w\) can no longer be interpreted as the Rabi frequency of an identifiable resonance. Physical time scales are instead related to the integrated decimated norm (also set by \(w\)), Eq.~\eqref{eqn:betaw_def}, rather than simply by \(t\sim1/w\).
\autoref{fig:Fig1}(b) illustrates resonance formation in the Jacobi algorithm, where arches are shown joining states that are involved in a resonance. The height of the arch indicates the frequency scale \(w\) associated with that resonance.

The full sequence of \(w\) values, not just the final eigenstates, encodes the scale-dependent information needed to quantify resonance proliferation. While the exact sequence is specific to the \(H\) in hand, one expects that coarse-grained statistical information describes properties of the entire ensemble of disorder realizations. 
The statistical analysis of the Jacobi algorithm has been coined the \emph{statistical Jacobi approximation} (SJA)~\cite{Long2023Fermi,hahn2025predicting}. Using the SJA, Ref.~\cite{Long2023Phenomenology} predicted that the proliferation of resonances would lead to stretched-exponential decay of autocorrelators in the thermalizing regime of disordered spin chains.

The SJA extracts the number of resonances with a frequency scale larger than \(w\), \(n_{\mathrm{res}}(w)\).
For models in the localized phase, new resonances eventually stop forming, so that \(n_{\mathrm{res}}(w)/N\) (the number of resonances per state) approaches a finite limit.
Models in the delocalized phase encounter more and more resonances as \(w\) decreases, such that \(n_{\mathrm{res}}(w)/N\) diverges as \(w \to 0\).
These resonances are also increasingly likely to involve states that have already been dressed by other resonances. 

If resonance formation at scale $w$ does not affect the rate of formation of subsequent resonances at scales less than $w$, the number of resonances admits a simple parametrization,
\(n_{\mathrm{res}}(w)/N \sim A\, w^{\theta} + \text{const}\).
Within this approximation, the \emph{resonance exponent} \(\theta\) distinguishes Hilbert-space-localized phases (\(\theta>0\), \(A<0\)) from delocalized ones (\(\theta<0\), \(A>0\)). 
This is confirmed within the toy model of a locally perturbed MBL chain~\cite{Crowley2022Constructive}, where the localization length of the chain determines the exponent \(\theta\) and local perturbations result in a proliferation of resonances above a critical localization length, \(\zeta_c\). The predicted \(\zeta_c\) turns out to agree with that predicted by the avalanche mechanism~\cite{DeRoek2017Stability}.

Resonance formation is expected to renormalize the bare localization length of the toy MBL model, and thus the exponent \(\theta\). We account for this renormalization by promoting the resonance exponent to a scale-dependent quantity, \(\theta=\theta(w)\).
This enables a frequency-resolved treatment of resonance proliferation that captures the crossover from the localized to the thermalizing regime.
Note that the Jacobi algorithm and SJA are not renormalization group (RG) procedures in the strict sense, as they do not eliminate degrees of freedom.

We present two derivations for two different flow equations for $\theta(w)$ based on the SJA (in Secs.~\ref{sec:HeuristicFlow} and ~\ref{sec:FunctionalFlow}). Both derivations ignore the correlations that build up between the matrix elements when running the Jacobi algorithm, and are valid in the regime where the matrix is sparse (which is characteristic of the localized regime with a few large off-diagonal elements connecting sites of the Hilbert space graph). Both predict a transition from a phase in which resonances do not proliferate (localized) to a proliferating phase (thermalizing). The onset of thermalization is signaled by an instability in the solution of the flow equations, i.e., a divergence of the resonance exponent, $\theta(w)\to -\infty$. Conversely, $\theta$ remains finite and positive at all frequencies in the localized phase, which is described by a line of fixed points $0<\theta(w=0)\leq 1$.

The two derivations differ in their assumptions about the form of \(n_{\mathrm{res}}(w)/N \) during the flow, and predict different universality classes for the transition between the localized and thermalizing phases. The first, in Sec.~\ref{sec:HeuristicFlow}, assumes a power-law form at all $w$, is simpler, produces Eq.~\eqref{eq:Heuristic}, and predicts a transition in the Berezinskii–Kosterlitz–Thouless (BKT)~\cite{1972JETP...34..610B,1973JPhC....6.1181K} universality class. 
The second, in Sec.~\ref{sec:FunctionalFlow}, allows for general \(n_{\mathrm{res}}(w)/N \), is more involved, and ultimately predicts that $\theta(w)$ diverges as a power law with a continuously variable exponent when a critical value is approached. We present the predictions of the two approaches, but leave open the nature of the transition.
The common and robust prediction of the two approaches is the existence and direction of the resonance-proliferation instability. Their different predictions for the critical scaling prevent us from assigning a universality class based on the present analysis, even if the numerical results are in closer agreement with the derivation in Sec.~\ref{sec:HeuristicFlow}.

Sec.~\ref{sec:numerics} demonstrates that the $\theta(w)$ predicted by the simpler flow equation~\eqref{eq:Heuristic} compares well to the numerically computed exponent in two models displaying Hilbert space localization, namely the Anderson model on regular random graphs (RRG), and a random matrix model with power-law distributed off-diagonal matrix elements. See~\autoref{fig:BoE_flow} and~\autoref{fig:fig3}.
In addition, we characterize how finite-size effects modify the resulting flow of $\theta$ (see~\autoref{fig:RRG_bump} and~\autoref{fig:RRG_N}).

We also show that the flow equation predicts qualitative features of the $\theta(w)$ vs $w$ curve in the standard model of MBL, namely the disordered Heisenberg model (\autoref{fig:fig3}(c) and \autoref{fig:XXZ_W7}).
Although this model is not Hilbert-space localized, the Jacobi flow still yields a numerically well-defined $\theta(w)$. At small disorder strengths, $\theta$ is initially negative, and diverges to larger negative values with decreasing $w$. This is the well-thermalizing regime. At intermediate disorder strengths, we observe that resonances are initially sparse, so that $\theta(w)>0$ for large $w$. However, the formation of a few resonances accelerates later resonance formation, which in turn causes $\theta(w)$ to become negative with decreasing $w$, and causes resonances to proliferate. We thus build a picture of slow thermalization in this regime through scale-dependent resonance formation and proliferation.

Although the exponent $\theta$ can be determined directly only within the Jacobi procedure, its flow controls physically relevant quantities such as eigenstate participation ratios and the return probability. In Sec.~\ref{sec:ObservablePredictions}, we argue that the number of resonances is a proxy for the participation ratio (see~\autoref{fig:nresvsPR}), while in~\autoref{fig:funct_retprob} we show that the prediction for the return probability obtained from the result [Eq.~\eqref{eq:solution_functional}] of the flow equation of Sec.~\ref{sec:FunctionalFlow} is in qualitative good agreement with the numerical results.

Our work ultimately sheds light on the mechanisms for resonance-driven crossovers and transitions.
We conjecture that the following models have resonance-driven Hilbert space localization transitions: the toy model of a locally perturbed MBL chain, the Anderson model on a random regular graph, and the L\'evy-Rosenzweig-Porter random matrix model (\autoref{sec:models}).
In particular, we predict that the toy MBL model with an initial \(\theta(w_0)>0\) may, as a result of a strong enough perturbation, actually flow to \(\theta(w) < 0\), and ultimately delocalize, so that large systems with bare localization lengths $\zeta < \zeta_c$ may thermalize at long times. 
We further conjecture that the crossover in prethermal MBL is also controlled by resonance proliferation, with observable features being described by an SJA flow, before being cut off by other effects such as avalanches, or by the finite system size.

More broadly, the SJA represents an important step towards a useful scale-dependent treatment of quantum dynamics, analogous to RG schemes.

\section{Resonances and the Jacobi Algorithm}
\label{sec:ResonancesSJA}

Both resonances and their proliferation, which will be the main objects of our study, can be given a precise meaning within the Jacobi algorithm, which we review in this section.
We first review the exact formulation of the algorithm used numerically (\autoref{subsec:JacobiAlgorithm}) and then describe the statistical Jacobi approximation (SJA) and the distributions which underlie our construction of the flow equations (\autoref{subsec:SJA}). The typical form of these distributions in the regime of interest is discussed in \autoref{subsec:SparseAnsatz}.

\subsection{Jacobi algorithm}
\label{subsec:JacobiAlgorithm}

Given an $N \times N$ Hamiltonian $H$, the Jacobi diagonalization algorithm iteratively diagonalizes the matrix of $H$ through a sequence of two-level (Givens) rotations~\cite{Jacobi1846}. We specialize our discussion to the case of a real symmetric \(H\).

Denote the matrix elements of \(H\) in an arbitrary computational basis $\{\ket{i_0} : i_0 = 1, \dots,N\}$ by \(H_{i_0 j_0} = \bra{i_0}H\ket{j_0}\). We denote by $w_0$ the largest off-diagonal matrix element (in modulus), i.e.,
\begin{equation}
    w_0 = \max_{i_0\neq j_0} |\bra{i_0}H\ket{j_0}| \equiv |\bra{a_0}H\ket{b_0}|,
\end{equation}
where \(\ket{a_0}\) and \(\ket{b_0}\) are the states which achieve this maximal matrix element.
The Jacobi algorithm performs a rotation between \(\ket{a_0}\) and \(\ket{b_0}\) such that the $2\times 2$ submatrix 
\begin{equation}
    H^{\mathrm{sub}} = 
    \begin{pmatrix}
        E_{a_0} & H_{a_0 b_0}\\
        H_{a_0 b_0} & E_{b_0}
    \end{pmatrix}
\end{equation}
(where $E_{i_0} = H_{i_0 i_0}$) becomes diagonal. This is accomplished with the rotation angle 
\begin{equation}\label{eqn:RotationAngle}
    \eta_0 = \arctan\frac{2 H_{a_0 b_0}}{E_{a_0}-E_{b_0}}
\end{equation}
and corresponding rotation \(R_0\) defined by
\begin{gather}
\label{eq:rotation_Jac}
    \ket{a_1} := R_0 \ket{a_0} = \cos \frac{\eta_0}{2} \ket{a_0} + \sin \frac{\eta_0}{2} \ket{b_0},\\
    \ket{b_1} := R_0 \ket{b_0} = \cos \frac{\eta_0}{2} \ket{b_0} - \sin \frac{\eta_0}{2} \ket{a_0},
\end{gather}
and \(R_0\ket{i_0} = \ket{i_0}\) for \(i_0 \not\in\{a_0,b_0\}\).
Defining new basis states \(\ket{i_1} = R_0 \ket{i_0}\), we have that \(H_{i_1 j_1}\) differs from \(H_{i_0 j_0}\) only in the two rows and two columns indexed by \(a_0\) and \(b_0\), and that \((R^\dagger H R)_{a_0 b_0} = 0\). 
This concludes one iteration within the Jacobi algorithm. See \autoref{fig:Fig1}(a).
We say that \(w_0\) has been decimated, in analogy to the terminology of the renormalization group.

After $n$ iterations of the algorithm, the matrix is expressed in the rotated basis
\begin{equation}
    \ket{i_n} = R_{n-1}\dots R_0 \ket{i_0}.
\end{equation}
It can be shown that the off-diagonal Frobenius norm 
\begin{equation}
    \norm{H}^2_{\text{off-diag}} = \frac{1}{N} \sum_{i \neq j} |H_{i_n j_n}|^2
\end{equation}
monotonically decreases, and converges to zero exponentially fast in the number of iterations with a rate that is at least \(1/N^2\). The decimated element \(w_n\) generally does not decrease monotonically with \(n\) but is numerically observed to decrease on average with fluctuations which are \(O(w_n)\) in size~\cite{Long2023Fermi}.

A particular rotation \(R_n\) corresponds to a resonance when the associated rotation angle \(\eta_n\) is larger than some threshold. Say
\begin{equation}
\label{eq:res_definition}
    |\eta_n| \geq \pi/4.
\end{equation}
In the localized regime, the dynamics can be characterized by the number of these resonances, as we make precise in the next section.

\subsection{Statistical Jacobi approximation}
\label{subsec:SJA}

The Jacobi algorithm exactly diagonalizes the Hamiltonian \(H\), so that exact analytic treatment must quickly become infeasible unless \(H\) is very simple.
The statistical Jacobi approximation (SJA) proposes that, rather than tracking the full information about which states are rotated in which order and with what angle, the dynamics of local observables should only require a statistical description of the Jacobi algorithm.
We thus now focus our attention on the distribution of decimated elements and identify two distinct regimes---sparse and dense---of the Jacobi flow [see Eqs.~\eqref{eq:SparseScaling}-\eqref{eqn:DenseScaling}]. In the sparse regime, we introduce the power-law exponent $\theta$, which will play a crucial role in the remainder of the paper. 

Following References~\cite{Long2023Phenomenology,Long2023Fermi,hahn2025predicting}, consider \emph{the distribution of decimated elements}
\begin{equation}\label{eqn:rhodec_def}
    \rhodec(w) = \sum_{n} 2 \delta(w - w_n),
\end{equation}
from which quantities, such as autocorrelation functions of certain observables and return probabilities, could be reconstructed~\cite{Long2023Phenomenology}.
This distribution encodes the density of decimated elements that occur in a small interval \((w-\rd w, w]\). The factor of \(2\) in Eq.~\eqref{eqn:rhodec_def} occurs because two matrix elements are decimated in each Jacobi rotation, one in the upper triangle and one in the lower triangle.
If \(w_n\) is monotonic in \(n\) (which is approximately the case~\cite{Long2023Fermi}), then the Jacobi index \(n(w)\) required to reach scale \(w\) can be recovered from \(\rhodec(w)\) as
\begin{equation}\label{eqn:ndec_integral}
    n(w) = \frac{1}{2}\int_w^\infty \rhodec(w') \,\rd w'.
\end{equation}

The distribution of decimated elements determines the density of resonances in the Jacobi algorithm as the scale \(w\) changes.
Modeling the energy denominators \(\omega_n = E_{a_n} - E_{b_n}\) encountered by Jacobi as being drawn independently from some continuous distribution \(p_\omega(\omega)\), the probability that a given rotation of scale \(w\) is a resonance is just the probability that \(|\omega| < w\), which is
\begin{equation}\label{eqn:ResonanceProbability}
    p_{\mathrm{res}}(w) = \int_{-w}^w p_\omega(\omega) \,\rd \omega \approx 2 p_\omega(0) w.
\end{equation}
The approximation in Eq.~\eqref{eqn:ResonanceProbability} holds for small \(w\), much less than the scale of variation of \(p_\omega(\omega)\), but still much larger than the level spacing.
The number density of resonances at scale \(w\) is thus
\begin{equation}\label{eqn:rhodec_res_relation}
    \rhores(w) = p_{\mathrm{res}}(w) \rhodec(w)/2 \approx p_\omega(0) w \rhodec(w).
\end{equation}
If \(w_n\) is monotonic in \(n\), then the total number of resonances encountered by the Jacobi algorithm before reaching scale \(w\) is
\begin{equation}\label{eqn:nres_def}
    n_{\mathrm{res}}(w) = \int_{w}^{\infty} \rhores(w') \,\rd w'.
\end{equation}

The distribution \(\rhodec(w)\), and hence the statistics of resonances, is determined by the evolution of the distribution of Hamiltonian matrix elements \(H_{i_n j_n}\) as the Jacobi algorithm proceeds.
This distribution is, after \(n\) Jacobi rotations,
\begin{equation}\label{eqn:rhoH_def}
    \rho_H(h | n) := \frac{1}{N} \sum_{i=1}^N \sum_{j \neq i} \delta(h - \bra{i_n}H\ket{j_n}).
\end{equation}
Here, \(N\) is the Hilbert space dimension, which is included so that \(\rho_H\) represents the average distribution of matrix elements in a row (or column) of \(H\). $\rho_H$ is normalized to \(N-1\). 

It is useful to distinguish two regimes based on the scaling of \(\rho_H\) with \(N\).
The \emph{sparse regime} is defined by
\begin{equation}
\label{eq:SparseScaling}
    \rho_H(h | n) = O(N^0)
    \quad\text{(sparse)}
\end{equation}
where \(h\) and \(n/N\) are fixed, and \(h\) is nonzero. In this regime, there are a few large \(O(1)\) matrix elements (\(O(1)\) per row), and many very small matrix elements. 

The opposite \emph{dense regime} is defined by the scaling
\begin{equation}\label{eqn:DenseScaling}
    \rho_H(h | n) = \tilde{\rho}_H(h \sqrt{N} | n/N^2)
    \quad\text{(dense)},
\end{equation}
where \(\tilde{\rho}_H\) is independent of \(N\). The \(N\) scaling in this formula is a consequence of all matrix elements being of similar size, together with the assumption that the Euclidean norm of a row in the Hamiltonian is \(O(1)\). In this case, we expect the bound on the convergence of the Jacobi algorithm (exponentially fast with rate \(1/N^2\)) to be saturated, leading to the rescaling of \(n \to n/N^2\).
A many-body Hamiltonian will also typically have \(\log N\) corrections to this scaling, as the Euclidean norm of a row will scale with the volume. We ignore these corrections.

Importantly, these sparse and dense \emph{regimes} refer to the scaling form of \(\rho_H\), not to whether \(H\) is sparse or dense as a matrix, in the sense of having most elements be strictly zero.

The sparse and dense regimes in the Jacobi algorithm indicate distinct dynamical behavior under Hamiltonian time evolution.
Specifically, the Jacobi algorithm probes characteristic spectral properties of localized and thermalizing systems.
Local Hamiltonians are sparse when expressed in a (quasi)local basis (that is, a basis related by finite-time evolution under a local Hamiltonian to a product state basis), since locality suppresses matrix elements between configurations that differ in many degrees of freedom. The Jacobi algorithm keeps the matrix in the sparse regime if the eigenstates are quasilocal in the starting product state basis.
By contrast, in systems satisfying the eigenstate thermalization hypothesis (ETH)~\cite{Deutsch1991quantum,Srednicki1994Chaos,Deutsch_2018_ETH}, the Hamiltonian expressed in the Jacobi basis resembles a random matrix at intermediate values of \(n\). Therefore, while the Hamiltonian eventually becomes diagonal, it does so by first becoming dense and then having all matrix elements decrease at similar rates.
Thus, a Hamiltonian that remains sparse asymptotically, down to the finite-size level-spacing cutoff, represents localization in the chosen basis. A finite intermediate sparse window may instead describe a prethermal regime: the system can appear localized over a range of scales before crossing into the dense regime at smaller \(w\). 

We aim to find a flow equation for \(\rho_H\) which is valid in the sparse regime. Thermalization will then be indicated by an instability of the flow equation (Secs.~\ref{sec:HeuristicFlow} and~\ref{sec:FunctionalFlow}).

An important note to the reader. The decimated element $w$ approximately monotonically decreases as the Jacobi algorithm is iterated, so that $w$ is a measure of the inverse flow time in the algorithm. All plots vs $w$ should therefore be read from right to left.

As we mentioned in the Introduction, the Jacobi scheme is not an RG procedure, as it does not eliminate degrees of freedom. Nevertheless, it shares similarities with known RG schemes.
For instance, the Jacobi construction is physically related to, but algorithmically distinct from, Aoki’s real-space decimation for Anderson localization~\cite{aoki1980,aoki1982}. Aoki decimation eliminates a site at a fixed external energy $E$, preserving the Green function at that energy and generating effective hoppings between the remaining sites. 
Jacobi instead performs a unitary diagonalization of the full Hamiltonian. Its rotation angle is determined by the instantaneous diagonal energies of the two states, and each rotation modifies two complete rows and columns. 
Jacobi therefore applies directly in many-body Hilbert space, but it does not produce a closed Aoki/Fisher-type real-space RG, even when applied to a one-dimensional single-particle problem. In both constructions, however, the generated couplings describe virtual hybridization processes, and a large Jacobi rotation signals the breakdown of perturbation theory when an off-diagonal matrix element becomes comparable to the corresponding energy mismatch.

The Jacobi method also differs from sparse implementations of strong-disorder RG, which discard sufficiently weak effective bonds~\cite{igloi2005,igloi2018,mard2017}, while the exact Jacobi numerics used in this work perform no such pruning. 
In the analytical SJA described in Sec.~\ref{sec:FunctionalFlow}, however, very small matrix elements are neglected, and thus effectively truncated. 
This resembles sparse SDRG in spirit, but the justification is different: we do not establish that the discarded terms are irrelevant at an infinite-randomness fixed point, but rather use it as an uncontrolled (but physically reasonable) approximation. 
An explicit sparse “Jacobi-Igloi” algorithm, in which generated matrix elements are retained according to a scale-dependent magnitude or resonance-probability cutoff, would constitute a distinct numerical method and is an interesting direction for future work.

\subsection{Ansatz for the sparse regime}
\label{subsec:SparseAnsatz}

In the sparse regime, it is natural to assume that the conditional distribution 
$\rho_H(h|n)$ develops a power-law tail at $|h| \ll w$~\cite{Crowley2022Constructive}.
This can be illustrated in the toy model of a locally-perturbed MBL spin-1/2 chain, described in \autoref{sec:Intro}. 

Recall that a spatially local perturbation $V$ in an ideal MBL chain generically has non-zero matrix elements between all the eigenstates of the chain~\cite{Huse2014Phenomenology}. However, as the eigenstates are quasi-local in a product state basis (or, as the eigenstates are labeled by a complete set of local integrals of motion (LIOMs)), the matrix elements are distributed according to a power law.
For a reference eigenstate with a given LIOM configuration, the number of basis states that differ from the reference only in a contiguous block of $r$ LIOMs grows as
\begin{equation}
    N(r) \sim 2^r .
\end{equation}
Quasilocality implies that typical matrix elements connecting the reference state to states differing within an \(r\)-sized block decay exponentially with \(r\),
\begin{equation}
    |V(r)| \approx w\, e^{-r/\zeta} \, 2^{-r/2}.
\end{equation}
The factor $2^{-r/2}$ ensures that the Frobenius norm over all $O(2^r)$ matrix elements at fixed range scales as $w e^{-r/\zeta}$, where $\zeta$ is a decay length.
Inverting the relation $|V(r^*)| \approx h$ gives
\begin{equation}
    r^*(h) \approx \frac{\log(w/h)}{\zeta^{-1} + \tfrac{1}{2}\log 2}.
\end{equation}
The number of matrix elements larger than $h$, therefore, scales as
\begin{equation}
    n(h) \approx \int_0^{r^*(h)} \rd r\, 2^r = O(2^{r^*(h)}) = O\left[\left(\frac{h}{w}\right)^{-\alpha} \right],
\end{equation}
with
\begin{equation}
    \alpha = \frac{\log 2}{\zeta^{-1} + \tfrac{1}{2}\log 2}.
\end{equation}
Differentiating with respect to $h$ yields
\begin{equation}
    \rho_H(h) = O(h^{-1-\alpha}),
\end{equation}
i.e., a power-law distribution at small $h$.

If we ignore the \emph{feedback effect} of Jacobi rotations on \(\rho_H\), namely the change in \(\rho_H\) due to the updates in the rows and columns containing $w_n$, then rotations simply decimate the elements of \(\rho_H\) one-by-one, and we get \(\rhodec(w) = \rho_H(w|0)\), so that \(\rhodec(w)\) is also a power law.
It is natural to parameterize the exponent as
\begin{equation}
    \rhodec(w) \propto w^{-2 + \theta}
\end{equation}
with \(\theta = 1-\alpha\).
Indeed from Eq.~\eqref{eqn:rhodec_res_relation}, this makes \(\rhores(w)\) another power law,
\begin{equation}
\label{eq:rho_res_w}
    \rhores(w) \propto w^{-1+\theta},
\end{equation}
such that the total number of resonances encountered in the algorithm, which is the integral of \(\rhores(w)\), will diverge for \(\theta \leq 0\) and remain finite for \(\theta >0\). (There is a finite system size cutoff at small \(w\) set by \(N\).) Notice that the extreme localized regime $\zeta=0$ gives $\theta = 1$.

One way to account for the feedback effect on \(\rho_H(h|n)\) is to promote \(\theta\) to be a function of \(w\).
In the subsequent sections, we will derive and analyze the flow equations predicting the behavior of $\theta(w)$.

\section{A heuristic flow equation for \texorpdfstring{\(\theta(w)\)}{theta(w)}}
\label{sec:HeuristicFlow}

\begin{figure}
    \centering
    \includegraphics[width=\linewidth]{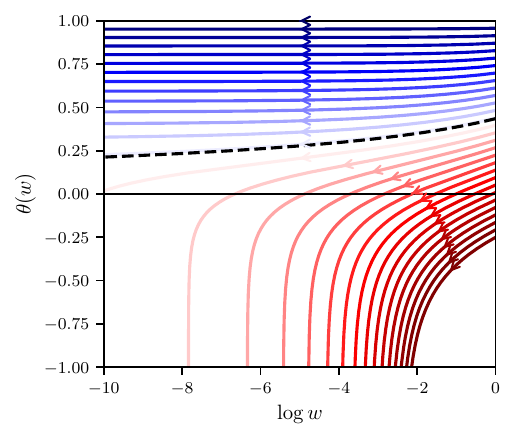}
    \caption{Numerical solution of the heuristic flow equations, Eq.~\eqref{eq:Heuristic}, for $f(\theta)=0.1(1-\theta)$. Different colors represent different initial conditions $(w=1, \theta_0)$, and the arrows indicate the direction of the flow. Blue curves represent the localized phase, while red curves represent the delocalized phase. The black dashed line is the separatrix, such that $\theta=0$ for $w=0$.}
    \label{fig:BoE_flow}
\end{figure}

We present the first derivation of a flow equation for the distribution of Hamiltonian matrix elements \(\rho_H\), and, specifically, the exponent \(\theta\), in the sparse regime of Jacobi.
The basic prediction of these heuristic flow equations, whose solution is given in Eq.~\eqref{eq:Heuristic} and illustrated in~\autoref{fig:BoE_flow} and~\autoref{fig:fig3}, is that the localized phase (\(\theta >0\)) is stable to sufficiently small perturbations in the Hamiltonian, but that there is an instability leading to a thermalizing phase (\(\theta \to - \infty\)) for large perturbations, or for arbitrarily small perturbations when \(\theta\) begins nonpositive.
Within the assumptions of this heuristic derivation, the flow has BKT-like topology and critical scaling. Let us make clear that throughout this work, “flow” refers to the scale dependence generated by successive Jacobi rotations. Neither the Jacobi algorithm nor the SJA eliminates degrees of freedom, so this flow is not an RG transformation in the usual sense.

We make several assumptions. 
First, we assume that the number of rotations \(n\) is parametrized by the current maximum matrix element \(w\). Indeed, exact numerics shows that the relationship between \(n\) and \(w\) is roughly monotonic, with small fluctuations~\cite{hahn2025predicting}. 

Second is a power law ansatz for \(\rho_H\), parameterized by the exponent \(\theta(w)\),
\begin{equation}
    \rho_H(h|\theta(w)) = C_{\theta(w)} |h|^{-2+\theta(w)} 1_{[h_{\min},w]}(|h|).
\end{equation}
Here, \(C_{\theta(w)}\) is a positive constant which depends on \(\theta(w)\), but not \(w\) directly, and the power law is supported on the double interval \(|h| \in [h_{\min},w]\), where \(h_{\min}\) is a function of \(N\). This ansatz is motivated by the sparse-regime counting argument discussed in \autoref{subsec:SparseAnsatz}, and thus specializes the flow equation to said sparse regime.
According to Eq.~\eqref{eq:rho_res_w}, the range of \(\theta\) is \(\theta \in (-\infty, 1]\), with \(\theta=1\) corresponding to a trivially localized model. 

The feedback of a Jacobi rotation on \(\rho_H\) is determined just by the angle \(\eta\) [see Eq.~\eqref{eq:rotation_Jac}]. When \(\eta\) is small, this feedback should be negligible, and thus our third assumption is that \(\theta\) only changes due to resonances, so that it can be expressed as a function of the number of resonances encountered \(n_{\mathrm{res}}\) [Eq.~\eqref{eqn:nres_def}]. This general form for the rate of change of \(\theta\) with the number of resonances per state, \(n_{\mathrm{res}}/N\), follows:
\begin{equation}
\label{eq:theta_res}
    \frac{\rd \theta(w)}{\rd (n_{\mathrm{res}}/N)} = - \tilde{f}[\theta(w)].
\end{equation}
Crucially, we assume that the right-hand side depends only on $\theta(w)$ and not separately on $w$. The rationale behind this assumption is that, once we have conditioned on the presence of a resonance, the specific value of $w$ at which it occurs becomes irrelevant. 
By assuming a resonance, we guarantee that \(\cos\tfrac{\eta}{2} \approx \sin\tfrac{\eta}{2} \approx \tfrac{1}{\sqrt{2}}\), and this value is what determines
the effect of the Jacobi rotation on the other matrix elements. The magnitude of \(w\) no longer enters.

The fourth and final assumption is that \(\tilde{f}(\theta)\) in Eq.~\eqref{eq:theta_res} is strictly positive for \(\theta <1\) and \(\tilde{f}(\theta=1) = 0\).
That is, resonances cause \(\theta\) to decrease, as they increase the tendency to thermalize.

These four assumptions lead to a highly constrained form for the evolution of \(\theta(w)\). 
Change variables in Eq.~\eqref{eq:theta_res} from \(n_{\mathrm{res}}\) to \(w\). If \(w\) is monotonic in \(n\), then \(\rd n_{\mathrm{res}}/\rd w = - \rho_{\mathrm{res}}(w)\) [from differentiating Eq.~\eqref{eqn:nres_def}]. 
Combining this with Eq.~\eqref{eq:rho_res_w} gives
\begin{equation}
\label{eq:res_w}
    \frac{\rd (n_{\mathrm{res}}/N)}{\rd w} = - C_{\theta(w)} w^{\theta(w)-1}.
\end{equation}
Combining Eqs.~\eqref{eq:theta_res} and \eqref{eq:res_w} gives the final form of the heuristic flow equation
\begin{equation}
\label{eq:Heuristic}
    \frac{\rd \theta(w)}{\rd w} = f[\theta(w)] w^{\theta(w)-1},
\end{equation}
where \(f[\theta(w)] = C_{\theta(w)} \tilde{f}[\theta(w)]\).
The solution to the flow equation (with a particular choice of \(f\)) is shown in \autoref{fig:BoE_flow}. 
(Recall that the number of resonances \(n_{\mathrm{res}}\) and the Jacobi index \(n\) increase from right to left in \autoref{fig:BoE_flow}.)

\autoref{fig:BoE_flow} exhibits a localized phase where \(\theta\) remains positive for all \(w\), and a thermalizing phase where \(\theta\) diverges to \(-\infty\) at a finite \(w_c\).
The existence of these two phases is generic for any strictly positive and continuous \(f(\theta)\).
Indeed, counting the number of resonances encountered late in the flow shows that it is possible to have an asymptote for \(\theta(w)\), \(\theta(w\to 0)\), which is strictly positive, but that whenever \(\theta(w)\) becomes nonpositive, it inevitably must diverge to \(-\infty\).
For an asymptote to exist, we need that
\begin{equation}\label{eqn:theta_change}
    \Delta \theta(w) = \int_0^{w} f[\theta(0)] h^{-1 + \theta(0)} \rd h \xrightarrow{w\to0} 0.
\end{equation}
This integral has an interpretation as being (proportional to) the number of resonances encountered in the final stages of the flow.
Eq.~\eqref{eqn:theta_change} holds for \(\theta(0)>0\), but the
integral diverges for any \(\theta(0) \leq 0\). As \(\theta(w)\) decreases as \(w\) decreases, this indicates that \(\theta(w)\) must eventually diverge if it ever hits \(\theta=0\).
This divergence signals the breakdown of the validity of the flow equation, which we interpret as a transition to the dense regime and eventual thermalization.

A separatrix terminating at \((0,0)\) separates the localized and thermalizing phases. The specific shape of the separatrix depends on \(f(\theta)\), but it typically approaches \(\theta=0\) very slowly as \(w\) is decreased. For instance, in \autoref{fig:BoE_flow}, when taking the initial power law to be \(\theta_0 = 0.2\), an initial perturbation magnitude of \(w_0 = e^{-10} \approx 4.5 \times 10^{-5}\) is already enough to land in the thermalizing phase. We observe that the localized regime is quantitatively not very stable near \(\theta=0\).

The phase diagram of \autoref{fig:BoE_flow} is highly reminiscent of the BKT flow equations. It has a line of stable points with  \(\theta>0\) and \(w=0\), and an unstable phase. Indeed, setting \(f(\theta)\) to a constant precisely reproduces the BKT RG flow equations~\cite{1973JPhC....6.1181K}. More generally, whenever \(f(\theta)\) is continuous and strictly positive near \(\theta=0\), the critical properties of the heuristic flow equations will be in the BKT universality class.

Existing RG analyses in several models exhibiting Hilbert space localization make the same predictions as Eq.~\eqref{eq:Heuristic}.
For instance, recent work~\cite{vanoni2023renormalization} on the Anderson model on expander graphs predicts a similar line of fixed points to that appearing in \autoref{fig:BoE_flow},
supporting the hypothesis that the transition to delocalization is driven by resonance proliferation, as captured by SJA.
The toy model of locally-perturbed MBL is also predicted to have a stable phase with \(\theta(0) > 0\)~\cite{Crowley2022Constructive}.

Curiously, while any asymptotic MBL transition should not be described by Hilbert space delocalization,
several RG analyses of the MBL transition have also predicted a BKT-like transition ~\cite{Goremykina2019Analytically,Dumitrescu2019Kosterlitz,Morningstar2019Renormalization,Niedda2025Renormalization,Morningstar2020Many}.
We expect that our Jacobi-based analysis can only capture resonance effects, but not the physics of rare regions in real space, which are crucial to these past RG analyses of MBL.
As such, it is not clear whether both analyses predicting BKT-like transitions is a coincidence, or if it indicates some (at least formal) connection between resonance physics and rare-region effects~\cite{Ha2023Resonances}.

\section{Application to models}
\label{sec:numerics}

\begin{figure*}
    \centering
    \includegraphics[width=\linewidth]{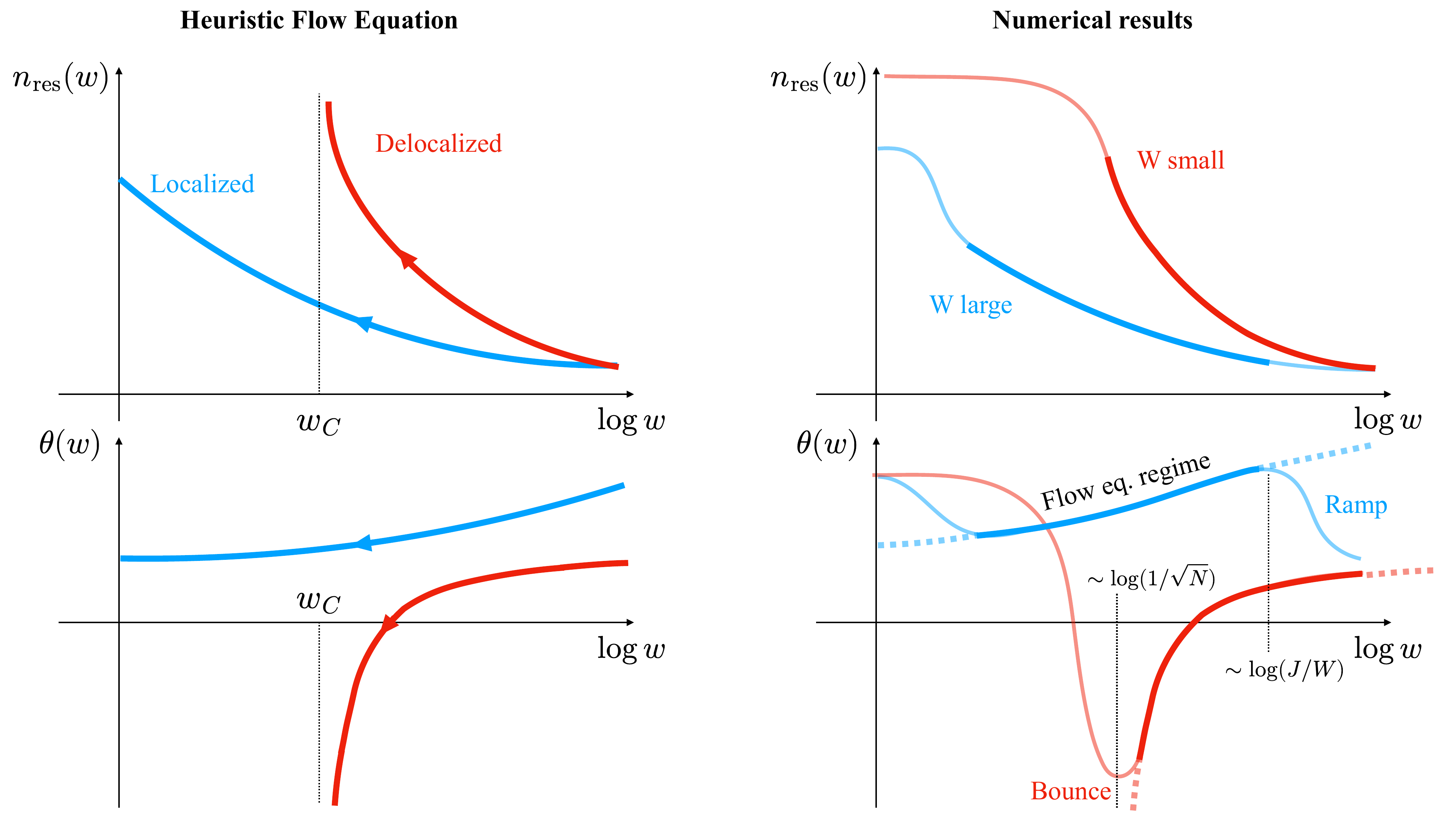}
    \caption{Sketch of the integrated number of resonances $n_{\mathrm{res}}(w)$ (top row) and of $\theta(w)$ (bottom row) as predicted by the heuristic flow equation (left) and found in the numerical simulations (right). The arrows indicate the direction of flow-time and thicker lines on the right panel indicate the regions where the numerical results are in agreement with the flow equation predictions. The heuristic flow equation predicts that, in the ergodic phase (red), $n(w)$ diverges at a finite value of $w=w_c$ and $\theta(w) \to -\infty$ as $w \to \infty$, while both quantities remain finite for all $\log (w)$ in the localized phase (blue). Numerically, the predicted divergences are not observed due to finite-size effects. Rather, $\theta(w)$ bounces back from an increasingly negative value and saturates to $\theta=1$. As discussed in the main text, the position of the bounce drifts with $N$ as $\log w_{\mathrm{bounce}} \sim \log(1/\sqrt{N})$. In models for which the Hamiltonian is a sparse matrix in the computational basis, a ramp is observed at large $w$ (small Jacobi index) and large $W$ (large diagonal disorder strength). As discussed in the main text, the ramp ends at $\log w_{\mathrm{ramp}} \sim \log(J/W)$.}
    \label{fig:fig3}
\end{figure*}

We compare the predictions of the heuristic flow equation to three different models of localization.
Namely: a random matrix model with power-law-distributed off-diagonal elements, the Anderson model on a random regular graph, and the widely studied disordered Heisenberg spin chain (\autoref{sec:models}).
For each of these models, we run the exact Jacobi diagonalization and compare the numerically observed \(\theta(w)\) to the prediction of the heuristic flow equations, finding qualitative agreement within the regime of validity of the flow equations (\autoref{sec:numerical_theta}). A sketch of the findings is reported in~\autoref{fig:fig3}.

Note that the SJA heuristic flow equation can only apply to the random matrix model and Anderson model asymptotically as $w\to 0$, as these models exhibit Hilbert space localization. That we also find qualitative agreement in the disordered Heisenberg spin chain indicates that the SJA flow equations can also describe prethermal MBL. 

\subsection{Three models of localization}
\label{sec:models}

\subsubsection{L\texorpdfstring{\'e}{e}vy-Rosenzweig-Porter}

As a first model, we consider a modification of the Rosenzweig-Porter (RP) random matrix ensemble~\cite{Kravtsov_2015_RandomMatrix}, with off-diagonal matrix elements extracted from a long-tailed L\'evy (power-law) distribution~\cite{Monthus_2017_Multifractal,kravtsov2020localizationtransitionrandomregular,Khaymovich2020Fragile,Biroli2021Levy,Kutlin2021Dynamical,Kutlin2024Anatomy,kutlin2024investigating,Lunkin2025Local}. The Hamiltonian $\HLevy$ of the model is given by
\begin{equation}
\label{eq:Ham_Levy}
    \HLevy = \mathcal{L}^{(\mu,\gamma)} + \mathcal{D},
\end{equation}
where $\mathcal{D}_{ij} = \epsilon_i \delta_{ij}$, $i=1,\dots,N$, is a diagonal matrix, with $\epsilon_i$ being independent and identically distributed (iid) random variables distributed uniformly in \([-1/2,1/2]\). 
The $N\times N$ off-diagonal random matrix $\mathcal{L}^{(\mu,\gamma)}$ has elements \(\mathcal{L}_{ij}\) distributed according to
\begin{equation}
\label{eq:distr_Levy}
    P_{\mu,\gamma}(\mathcal{L}_{ij}) = \frac{\mu}{2 N^{\gamma} \mathcal{L}_{ij}^{1+\mu}} \mathbf{1}_{[N^{-\gamma/\mu},\infty)}(|\mathcal{L}_{ij}|),
\end{equation}
where \(\mathbf{1}_A\) is the indicator function for the set \(A\) and $\gamma$ and $\mu$ are parameters. In particular, $1+\mu$ is the exponent of the power-law distribution, while $\gamma$ governs the magnitude of the off-diagonal elements. The typical largest element across rows or columns of $\mathcal{L}_{\mu,\gamma}$ is of order $N^{(1-\gamma)/\mu}$, so that choosing $\gamma=1$ sets the typical largest element to be \(O(N^0)\).
[The typical largest element in the entire matrix will still be \(O(N^{1/\mu})\) with this choice.]

The phase diagram of the Lévy-RP (LRP) model is described in Ref.~\cite{Biroli2021Levy}.
In particular, for $\mu<1$ and $\gamma>1$, the spectrum is localized (except for a vanishing fraction of states in the Lifshitz tails).
That is, the eigenstates of the model have a finite participation ratio in the computational basis (that is, the basis used in Eq.~\eqref{eq:distr_Levy}) as \(N\) increases, so that they are essentially supported on only a few basis states.
Indeed, in this regime, the typical largest element per row is asymptotically smaller than the diagonal terms, so that the off-diagonal matrix elements can mostly be treated perturbatively.
For $\mu<1$ and $\gamma<1$, a mobility edge separates ergodic states, having $|E|< N^{(1-\gamma)/\mu}$, from localized states, having $|E|> N^{(1-\gamma)/\mu}$.

Our primary interest is the line $\gamma = 1$. 
This line is marginal, as the diagonal part of \(\HLevy\) has the same \(O(1)\) scaling with \(N\) as the typical largest element of the off-diagonal part.
The off-diagonal matrix elements can thus compete with the diagonal energy difference, and the behavior of the system is determined by $\mu$.
We expect that there is a critical \(\mu_c\) such that \(\HLevy\) is fully localized for \(\mu < \mu_c\) and delocalized, with possibly a mobility edge, for \(\HLevy > \mu_c\).
For $1 < \mu < 2$ and for both $\gamma<1$ and $\gamma>1$, the model is delocalized~\cite{Biroli2021Levy} (ergodic and multifractal, respectively), so we expect that \(\mu_c < 1\).

\subsubsection{Random Regular Graph}

As a second model, we consider the Anderson model on a random regular graph (RRG), with Hamiltonian
\begin{equation}
    \HRRG = \mathcal{A} + \mathcal{D}.
\end{equation}
The diagonal matrix $\mathcal{D}$ has the same definition as for the Levy RP model, while $\mathcal{A}$ is the adjacency matrix of a graph uniformly sampled from the set of all graphs with \(N\) vertices and fixed degree. That is, the adjacency matrix of an RRG. RRGs are, with high probability, expander graphs, so that they are locally tree-like. Loops, which are necessary to preserve the regularity, are typically large-scale. 
The interest in Anderson localization on expander graphs is motivated by the work of Ref.~\cite{Basko06}, in which the problem of real-space MBL was mapped to that of localization on an associated configuration graph, which is locally tree-like. Both problems have regimes of slow dynamics and exhibit large finite-size drifts of the crossovers between delocalized and localized phases~\cite{sierant2023universality}. Nevertheless, the Anderson problem on RRGs does not capture important real-space correlations present in MBL~\cite{tikhonov2021AndersonMBL}.

It is widely accepted that the Anderson model on RRGs has a localization transition at $W_c = 18.1 \pm 0.1$ (when the degree is three)~\cite{tikhonov2016Anderson,Lemarie2022critical,sierant2023universality,vanoni2023renormalization}.
For $W<W_c$, the states at the middle of the band are ergodic and satisfy ETH~\cite{sierant2023universality,vanoni2023renormalization,vanoni2023analysis}, while for disorder strengths $W > W_c$, the system becomes fully localized.
The localized phase is described by a line of fixed points parametrized by $W$~\cite{vanoni2023renormalization}. We do not attempt to extract \(W_c\) from the Jacobi data presented below, but take this established value as an input when discussing the RRG phase diagram.

The SJA and the cavity Green-function approach provide complementary descriptions of localization on the RRG. The running function \(\theta(w)\) characterizes the distribution of Jacobi-decimated matrix elements across scales and hence the scale dependence of resonance formation. Through the associated resonance distribution \(\rho_{\mathrm{res}}(w)\), its integrals and transforms give access to quantities such as participation ratios and return probabilities, as discussed in Sec.~\ref{sec:ObservablePredictions}. It remains open whether further multiscale observables, including LDOS statistics and multifractal properties, can be extracted from this flow. A possible extension would retain a joint distribution \(\rho_H(h,r|w)\), where \(r\) resolves the graph distance associated with an effective matrix element.

\subsubsection{XXZ spin chain}
\label{sec:models_XXZ}

As a third model, we consider the random field spin-\(1/2\) XXZ chain at the Heisenberg point, often called the standard model of MBL, defined by the Hamiltonian
\begin{equation}
    \HXXZ = J\sum_{i=1}^{L}\left(\hat S_i^x\hat S_{i+1}^x+\hat S_i^y\hat S_{i+1}^y+\Delta \,\hat S_i^z\hat S_{i+1}^z\right)+\sum_{i=1}^{L}h_i\hat S_i^z,
\end{equation}
where the random fields $h_i$ are independently and uniformly distributed in $[-W,W]$. In the main text, we study the isotropic point $\Delta=1$, corresponding to the random-field Heisenberg chain. The noninteracting limit $\Delta=0$ maps, through the Jordan-Wigner transformation, to an Anderson-localized chain of free fermions. In Appendix~\ref{app:MBL_Anderson}, we use the many-body eigenbasis of this $\Delta=0$ Hamiltonian as an alternative starting basis for the Jacobi algorithm applied at $\Delta=1$.

The presence of a putative MBL phase in the XXZ disordered spin chain was put forward in Ref.~\cite{Znidaric2008manybody}, and has received much attention since then (see Ref.~\cite{sierant24MBLreview} for a recent review).
Note that localization in this model refers to localization in real space, rather than the Hilbert space localization of our other two models.
Consequently, a positive \(\theta\) over a finite interval in the XXZ chain should not be interpreted as evidence for asymptotic Hilbert-space localization. Here, the Jacobi flow is used to characterize the formation of many-body resonances over the finite frequency and system-size window accessible numerically.
With increasing system sizes, various spectral measures drift towards their values in the thermal phase, so that there is currently no consensus as to whether or not MBL is stable~\cite{vsuntajs2020quantum,Sierant20Thouless,Sels2021Dynamical,abanin2021distinguishing,panda2020can}.
Quantitative analysis of the so-called avalanche instability places the critical disorder strength for a localization transition above \(W_c \gtrsim 20\)~\cite{Morningstar2022Avalanches,Sels2022avalanches}, with the value being set by current numerical capabilities.

However, at more moderate values of disorder strength, the short-to-intermediate time dynamics which can be observed in quantum simulator experiments and numerics has been proposed to be governed by many-body resonances~\cite{Gopala2019Instability,Crowley2022Constructive,Long2023Phenomenology,Morningstar2022Avalanches,Ha2023Resonances}. It is this regime that we can hope to capture with the SJA.

\subsection{Numerical comparison to the flow equations}
\label{sec:numerical_theta}

For each of the models discussed above, we apply the Jacobi algorithm to their Hamiltonian and compare $\theta(w)$ to the prediction of the flow equations. We find qualitative agreement between the numerics and the predictions of the heuristic flow equations, within their regime of validity.

\begin{figure*}
    \centering
    \includegraphics[height=0.38\textwidth]{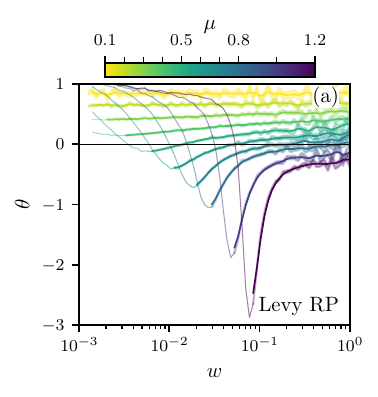}
    \includegraphics[height=0.38\textwidth]{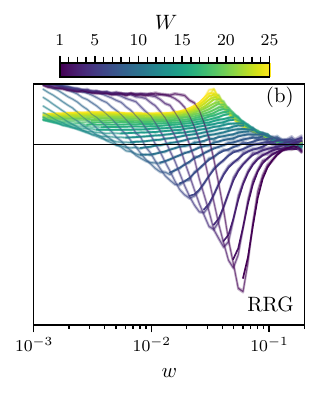}
    \includegraphics[height=0.38\textwidth]{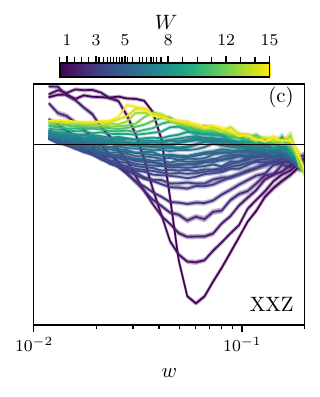}
    \caption{Jacobi flow of $\theta(w)$ in the three models presented in Sec.~\ref{sec:models}. The solid thin lines are the bare numerical data, and the shaded area around the curves represents the statistical error on $\theta$, obtained using the bootstrapping method applied to the distribution of decimated elements and propagated to $\theta$. The solid thick lines are obtained by applying a weak Gaussian filter to the numerical data to remove the fluctuations at large values of $w$; the thick lines are plotted until $\theta$ reaches its minimum value, i.e., before the flow enters the dense regime (see text and~\autoref{fig:fig3}). The ticks on the color bars are positioned at the values of the parameter of the displayed curves.
    The panels show $\theta(w)$ for (a) the Levy-RP model, defined in Eq.~\eqref{eq:Ham_Levy}, with $N=1024$ and averaged over $O(10^4)$ disorder realizations,
    (b) the Anderson model on RRG, with connectivity three, $N=2^{L}$, $L=10$, and averaged over $10000$ realizations, and 
    (c) the XXZ spin chain with $L=14$, in the total magnetization sector $\sum_{i=1}^L \hat{S}_i^z = 0$, and averaged over $3000$ realizations. All three cases qualitatively agree with the prediction of the heuristic flow equation Eq.~\eqref{eq:Heuristic}, see~\autoref{fig:BoE_flow} and~\autoref{fig:fig3}. 
    }
    \label{fig:theta_flow}
\end{figure*}

The exponent $\theta(w)$ can be numerically extracted by taking the discrete derivative of the distribution of decimated elements.
In particular:
\begin{equation}
    \theta(w) = 1 + \frac{\rd \log \rhodec(\log w)}{\rd \log w},
\end{equation}
where \(\rhodec(\log w) = w \rhodec(w)\).
The numerical procedure tracks the number of off-diagonal elements that have been decimated in an interval $(w/w_{\mathrm{step}},w]$ (with $w_{\mathrm{step}} = 1.1$) for each realization of a Hamiltonian $H$. After averaging over many disorder realizations, the distribution $\rhodec(\log w)$ is obtained from the histogram of the number of decimated elements in each logarithmic bin.
$\theta(w)$ is then computed as a finite difference
\begin{equation}
\label{eq:theta_num}
    \theta(w) \approx 1+\frac{\Delta \log \rhodec(\log w)}{ \log w_{\mathrm{step}}},
\end{equation}
where $\Delta f(x) = f(x+\Delta x) - f(x)$.

We present the numerical results for the LRP model, the Anderson model on an RRG, and the XXZ spin chain in \autoref{fig:theta_flow}. The following subsections explain the important features of these plots and how they should be compared to \autoref{fig:BoE_flow}. 
We explain the salient features of \autoref{fig:theta_flow}, schematically reported in~\autoref{fig:fig3}, starting at large \(w\) (early in the Jacobi flow) and moving towards smaller \(w\) (later in the Jacobi flow). All models qualitatively reproduce the predictions of the heuristic flow equation within its regime of validity. In the case of the XXZ spin chain, the same features are reproduced by initializing the Jacobi algorithm in the non-interacting $\Delta=0$ basis, rather than the $S^z$ product-state basis, as shown in Appendix~\ref{app:MBL_Anderson}.

\subsubsection{Ramp}

The monotonic increase of the exponent $\theta(w)$ as $w$ decreases from its starting value is dubbed ``the ramp" (right panel of~\autoref{fig:fig3}). 
This feature is visible only in the large-\(w\)  and large-disorder regime, and appears in both the RRG and XXZ models.

The ramp is not captured by the heuristic flow equation because the off-diagonal matrix elements of the Hamiltonian do not satisfy the power-law ansatz in the starting basis (defined by the basis in which disorder is diagonal).
In this basis, the Hamiltonian only contains a finite number of $O(1)$ matrix elements per row, while all remaining entries are strictly zero.
Several Jacobi rotations are needed to set up a power-law distribution in these models, and this initialization process results in the ramp.

A heuristic to estimate the value of \(w\) at which the power law in the RRG and XXZ models should emerge, and thus where the ramp ends, can be made as follows. 
In order for the power law distribution to emerge, the Jacobi algorithm must at least decimate all the initially large bare matrix elements in the Hamiltonian.
At large $W$, resonances are rare, and therefore the small-angle approximation is valid $\eta_n \sim 2w/W \ll 1$, so that newly generated matrix elements in Eq.~\eqref{eq:rotation_Jac} will be a factor of \(1/W\) smaller than the initial matrix elements. 
Thus, once $w \approx 1/W$, the vast majority of the initial matrix elements in the Hamiltonian have been decimated, and a power-law distribution of elements has the opportunity to emerge.

\autoref{fig:RRG_bump} confirms that the ramp stops at $w \approx 1/W$ for large values of $W$ in the RRG.
Thus, only the regime $w < 1/W$ could be captured by the flow equation~\eqref{eq:Heuristic} in the RRG and XXZ models. \autoref{fig:RRG_bump} is not a finite-size estimate of a localization threshold. It tests only the initialization scale \(w_{\mathrm{ramp}}\sim1/W\), above which the bare sparsity of the Hamiltonian controls the flow. The new information supplied by the Jacobi analysis is the scale-resolved structure and proliferation of resonances, rather than a new determination of \(W_c\).

\subsubsection{Flow equation regime}

Once the power law distribution of Hamiltonian matrix elements has been initialized, there is an intermediate range of \(w\) which is well described by the heuristic flow equations.

In particular, starting from larger $w$, all plots show a flow towards more negative \(\theta\) with decreasing $w$.
Models which are more strongly disordered (larger \(W\) or smaller \(\mu\)) begin at a larger, positive value of \(\theta\), and their rate of flow towards negative values is more gradual. These curves may saturate at a nonzero positive \(\theta\).
On the other hand, less disordered models begin with a negative \(\theta\), which rapidly flows to become even more negative.

\begin{figure}
    \centering
    \includegraphics[width=\linewidth]{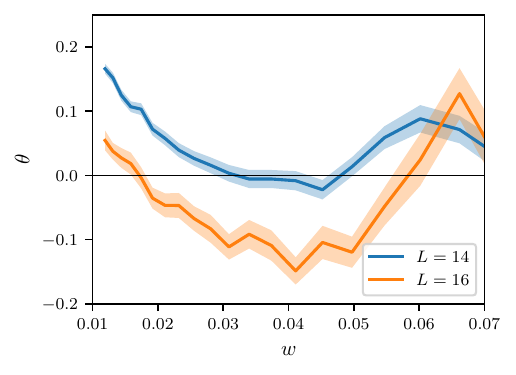}
    \caption{The exponent $\theta(w)$ in the XXZ spin chain at disorder strength $W=7$ zoomed into the range of $w$ where $\theta \approx 0$. In each curve, the solid line represents a windowed average, while the shaded area represents the error, defined as one standard deviation from the mean. The p-value for having $\theta >0$ at the maximum of the curve is $p_> = 0.997$ for $L=16$, $p_> = 1$ for $L=14$, while the p-value for having $\theta <0$ at the minimum of the curve is $p_<=1$ for $L=16$, $p_<=0.93$ for $L=14$. These results show that in the XXZ spin chain, there are values of the initial parameters for which $\theta(w)$ decreases from positive to negative values as $w$ decreases, which we interpret as resonance proliferation. }
    \label{fig:XXZ_W7}
\end{figure}

\begin{figure}
    \centering
    \includegraphics[width=\linewidth]{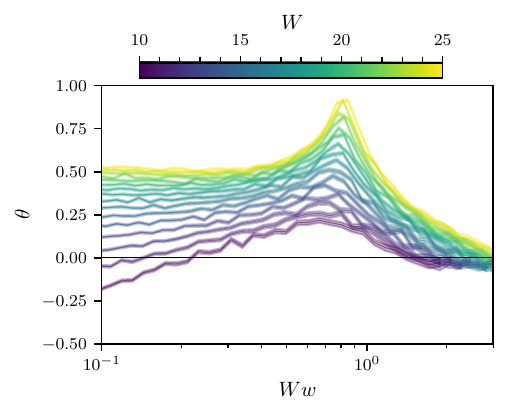}
    \caption{The exponent $\theta(w)$ in the RRG for large values of the diagonal disorder strength $W$. When rescaling the $x$-axis by $W$, the maxima of the ramps present in~\autoref{fig:theta_flow}(b) align. This collapse concerns the end of the initial ramp and should not be interpreted as an estimate of \(W_c\).}
    \label{fig:RRG_bump}
\end{figure}

Remarkably, in all the models, there are initial conditions for which $\theta>0$ at large $w$ and the Jacobi flow drives it to negative values, $\theta<0$, at smaller \(w\).
This signals the presence of resonance proliferation at intermediate disorder strengths, and is a key prediction of the flow equations.
While for the LRP and RRG models, this effect is clearly visible in~\autoref{fig:theta_flow}, for the XXZ spin chain, the effect is milder, and can be better observed in~\autoref{fig:XXZ_W7}, where the curves for $W=7$ and $L=14$ and $L=16$ are shown at a closer scale, displaying the aforementioned transition from positive to negative $\theta$.
\autoref{fig:XXZ_W7} does not locate an MBL transition or determine the asymptotic fate of the chain. It shows that, at the accessible sizes, there is a range of parameters for which the Jacobi flow begins in a sparse, localized-like regime and subsequently crosses to \(\theta<0\). The contribution of the SJA is to identify the buildup and proliferation of resonances underlying this finite-size crossover and to organize it by the frequency scale \(w\).

\subsubsection{Bounce}
\label{Sec:Bounce}
Once \(w\) becomes very small, $O(1/\sqrt{N})$, \autoref{fig:theta_flow} departs from the predictions of the flow equation. The asymptotic divergence (\(\theta<0\)) or saturation (\(\theta>0\)) of \(\theta(w)\) is interrupted by a bounce back to \(\theta = 1\). The bounce occurs in correspondence with the ``knee'' in \(n_{\mathrm{res}}\) [see~\autoref{fig:fig3}].

The bounce is a signature of the models exiting the sparse regime and crossing over to the dense regime, where Eq.~\eqref{eq:Heuristic} no longer applies.
Indeed, we note that the value \(\theta=1\) is characteristic of the dense regime. In Appendix~\ref{app:GOE}, we demonstrate that a typical random matrix from the Gaussian orthogonal ensemble (GOE) has \(\theta(w) \approx 1\) for all \(w\).
In addition, the position of the bounce in \(w\), and how this changes with the Hilbert space dimension \(N\), also suggests that this feature is associated with the dense regime. \autoref{fig:RRG_N} shows that the value of $w$ at the bounce position decreases like \(1/\sqrt{N}\), which is the typical system-size dependence of matrix elements in the dense regime.

\subsubsection{Scaling with matrix size}
\label{sec:N_scaling}

Lastly, we comment on how \(\theta(w)\) depends on the matrix size, via the Hilbert space dimension \(N\).

\begin{figure}
    \centering
    \includegraphics[width=\linewidth]{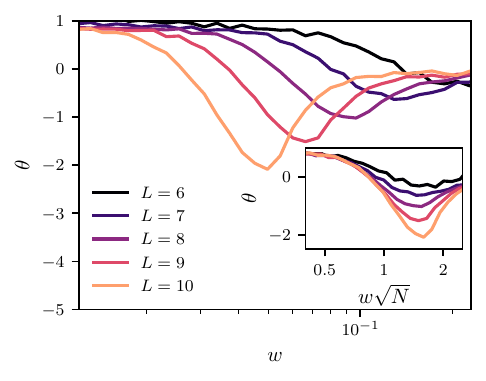}
    \caption{The exponent $\theta(w)$ in the RRG for $W=2$ and different system sizes $N=2^L$. When rescaling the x-axis by $\sqrt{N}$ (see inset), the minima align, showing that the Jacobi flow is in the dense regime when $\theta(w)$ is minimal.
    }
    \label{fig:RRG_N}
\end{figure}

The flow equations do not directly predict the $N$ dependence for \(\theta(w)\). 
The SJA analysis indicates that the flow equation Eq.~\eqref{eq:Heuristic} should be independent of $N$ in the sparse regime.\footnote{There are effects that could allow the unspecified function \(f(\theta)\) appearing in Eq.~\eqref{eq:Heuristic} to have a weak system size dependence. For instance, in the XXZ model, the norm of a row in the Hamiltonian scales with \(\log N\), so we expect \(\log N\) effects at least in that model.}
However, the sparse regime scaling breaks down when \(\theta(w)\) begins to diverge, and the dense regime emerges. Thus, there must be a different system size dependence near this divergence.
In particular, the value of \(w\) at the bounce decreases as \(1/\sqrt{N}\) (as remarked in Sec.~\ref{Sec:Bounce}), and the depth of the minimum in \(\theta(w)\) grows with \(N\)
(\autoref{fig:RRG_N}).
Thus, while the prediction of the heuristic flow equation that \(\theta(w)\) diverges at a fixed finite \(w_c\) is not reliable, there is a developing signature of the putative divergence at finite \(N\).
However, this divergence does not occur at a fixed \(w_c\).

Note that the emergence of the dense regime at smaller values \(w < w_c \sim 1/\sqrt{N}\) as \(N\) is increased does not imply that thermalization occurs at later real times \(t\). The correct conversion from decimated elements \(w\) to time \(t\) is to compare \(t\) to~\cite{Long2023Phenomenology,Long2023Fermi}
\begin{equation}\label{eqn:betaw_def}
    \beta(w) = \left( \int_{w}^{w_0} w'^2 \frac{\rhodec(w')}{N} \,\rd w \right)^{-1/2}.
\end{equation}
In the dense regime, matrix elements \(w\) scale as $1/\sqrt{N}$, but \(\rhodec\) is proportional to $N^2$, so that the integral above is \(O(1)\).

\section{Systematic flow equations in the sparse regime}
\label{sec:FunctionalFlow}

The heuristic equation of Sec. III reproduces the observed direction of the Jacobi flow but assumes a particular response of \(\theta\) to each resonance. In this section, we derive a functional evolution equation for the full distribution \(\rho_H(h|n)\). The main result is Eq.~\eqref{eq:solution_functional}: the solution again has a stable sparse regime with \(\theta>0\) and an instability when \(\theta\) becomes negative, but it does not reproduce the BKT-like critical scaling of the heuristic flow equation. The technical solution of the functional equation is given in Appendix~\ref{app:FunctionalFlowSolutionAppendix}.

More specifically, we obtain a nonlinear integrodifferential flow equation Eq.~\eqref{eqn:FullFlow} for the distribution \(\rho_H(h|n)\) (\autoref{sec:FunctionalFlowConstruction}). The derivation assumes that the magnitude of a matrix element is uncorrelated with its position in the matrix of \(H\). This is the SJA, which treats matrix elements and the Jacobi rotations acting on them statistically. This assumption should be valid within sufficiently small energy windows of the Hamiltonian, which may still contain $O(N)$ states.

However, Eq.~\eqref{eqn:FullFlow} is too complicated to make further progress, and approximations are required to analyze it effectively.
We present two levels of approximation, the first being more aggressive than the second.

The first level (\autoref{sec:VerySparseRegime}) assumes that \(H\) is sufficiently sparse so that Jacobi rotations between two simultaneously large matrix elements can be neglected. In a rotation \(h_1 \mapsto \cos(\eta/2)\, h_1 + \sin(\eta/2)\, h_2,\) we therefore treat one of the two elements \(h_{1,2}\) as negligible, since it is typically much smaller than the other. This approximation yields Eq.~\eqref{eqn:ApproxFlowq}, which admits the analytical solution reported in Eq.~\eqref{eq:theta_funct1}.
However, this solution fails to reproduce the resonance-proliferation instability: the first approximation predicts that \(\theta(w)\) always flows to \(\theta \to 1\), corresponding to the strongly localized regime.

The second level of approximation (\autoref{sec:FunctionalFlowInstability}) models terms that were neglected in Eq.~\eqref{eqn:ApproxFlowq} in a form that still allows for a controlled analysis and does exhibit a resonance-proliferation instability.
Specifically the approximation accounts for rotations for which \(\cos(\eta/2)\, h_1 + \sin(\eta/2)\, h_2 > h_1\) by hand.
A perturbative solution of the new functional equation, Eq.~\eqref{eq:solution_functional}, qualitatively reproduces the behavior of the heuristic flow equation, as shown in~\autoref{fig:sparse_theta}. In particular, it has a line of fixed points with \(\theta>0\), and a separatrix at \(\theta = 0\), below which the solution to the flow equation must eventually diverge. Nonetheless, some quantitative features of the heuristic flow equation remain absent, such as the specific BKT critical exponents. 

Let us clarify that this approximate treatment of matrix-element distributions is not known to become asymptotically exact and is therefore an uncontrolled approximation. It is most justified deep in a sparse localized regime, where most rotations are weak, and the correlations generated by any individual rotation affect only a small statistical fraction of matrix elements. It may fail when resonances are strongly correlated in Hilbert space, as can occur when the Hilbert-space couplings inherit real-space locality. Establishing a controlled parameter for this approximation remains an open problem.

\subsection{Functional flow equation}
\label{sec:FunctionalFlowConstruction}

Within the SJA, a functional flow equation for the distribution 
\(\rho_H(h|n)\) can be obtained. Namely, we assume independence of the two 
matrix elements, \(h_1\) and \(h_2\), affected by the Jacobi rotations, so 
that they are each sampled from \(\rho_H(h|n)\) itself. With this assumption, and neglecting subleading in $N$ contributions, we derive the flow equation Eq.~\eqref{eqn:FullFlow}.

Introduce $\mathrm{w}_n = \bra{a_n}H\ket{b_n}$, that is, the current maximal off-diagonal element in modulus taken with its sign (so $w_n = |\mathrm{w}_n|$).
The integral of \(\rho_H(h|n)\) with respect to \(h\) is \(N-1\) if \(H\) is \(N \times N\) [see Eq.~\eqref{eqn:rhoH_def}]. For convenience, we consider \(H\) which is \((N+1) \times (N+1)\), so that \(\rho_H(h|n)\) integrates to \(N\).\footnote{Regardless, this results in subleading-in-\(N\) corrections.}

A single Jacobi rotation decimates two elements at \(h = \mathrm{w}_n\), replacing them with \(0\), and updates \(4(N-1)\) off-diagonal elements.
The corresponding update to the distribution is
\begin{widetext}
\begin{multline}\label{eqn:FullFlow}
    \rho'_H(h) - \rho_H(h) = \frac{2}{N+1}\left[\delta(h) - \delta(h-\mathrm{w}_n)\right] 
    + 2\frac{N-1}{N+1} \bigg\{\int \big[\delta(h - c_n h_1 - s_n h_2) - \delta(h-h_1) \\
    + \delta(h - c_n h_2 + s_n h_1) - \delta(h-h_2)\big] \frac{\rho_H(h_1)}{N} \frac{\rho_H(h_2)}{N} \rd h_1 \rd h_2 \bigg\}.
\end{multline}
For brevity, we denote \(\rho_H(h) = \rho_H(h|n)\), \(\rho'_H(h) = \rho_H(h|n+1)\), \(c_n = \cos(\eta_n/2)\), and \(s_n = \sin(\eta_n/2)\).
The first term on the RHS of the equation accounts for the decimation at \(h=\mathrm{w}_n\). The integral replaces \(2(N-1)\) randomly sampled elements \(h_1\) with \(c_n h_1 + s_n h_2\) and another \(2(N-1)\) elements \(h_2\) with \(c_n h_2 - s_n h_1\) in the entire matrix.

One can verify that the update equation Eq.~\eqref{eqn:FullFlow} preserves the normalization of \(\rho_H(h)\) and reduces the off-diagonal Frobenius norm of \(H\) [given by the integral of \((N+1) h^2 \rho_H(h)\)] by exactly \(2 w_n^2\).
For later convenience, we also note that some of the delta function integrals can be performed explicitly,
\begin{equation}
\label{eq:rhoNsquared}
    \int [\delta(h-h_1) + \delta(h-h_2)] \frac{\rho_H(h_1)}{N} \frac{\rho_H(h_2)}{N} \rd h_1 \rd h_2
    = 2\frac{\rho_H(h)}{N}.
\end{equation}

\subsection{First approximation: very sparse regime}
\label{sec:VerySparseRegime}

Equation~\eqref{eqn:FullFlow} is complicated by the nonlinear \(\rho_H(h_1) \rho_H(h_2)/N^2\) term.
In the sparse regime, the effect of this term should be subleading in \(N\), as we quantify in this section.
Our first level of approximation is to neglect this subleading term. This results in a simpler flow equation [Eq.~\eqref{eqn:ApproxFlowq}], which admits an analytical solution for \(\theta(w)\) [Eq.~\eqref{eq:theta_funct1}]. However, this solution does not exhibit the instability to resonance proliferation, indicating that some of the effects we neglected by dropping the subleading term are necessary for thermalization.

In the sparse regime, \(\rho_H(h)\) can be approximated as 
\begin{equation}\label{eqn:SparseFormRho}
    \rho_H(h) = (N - m) \delta(h) + q_H(h),
\end{equation}
where
\begin{equation}
    \int_{-\infty}^\infty q_H(h) \rd h = m = O(1).
\end{equation}
That is, most matrix elements are very small, and there are $m = O(1)$ non-zero elements per row. The approximation should be viewed as imposing some lower cutoff on the size of the matrix elements we consider to be important, and setting all elements smaller than that to zero.
Substituting Eq.~\eqref{eqn:SparseFormRho} into Eq.~\eqref{eqn:FullFlow} and using Eq.~\eqref{eq:rhoNsquared} provides update equations for \(q_H\),
\begin{multline}\label{eqn:FullFlowq}
    q'_H(h) - q_H(h) = -\frac{2}{N+1}\delta(h-\mathrm{w}_n) - 4 \frac{N-1}{N+1} \frac{q_H(h)}{N} 
    + 2\frac{N-1}{N+1}\frac{N-m}{N^2} \bigg\{ \frac{2}{c_n} q_H\left( \frac{h}{c_n}\right) + \frac{1}{|s_n|}\left[q_H\left( \frac{h}{s_n}\right) + q_H\left(-\frac{h}{s_n}\right) \right] 
    \bigg\} \\
    + \frac{2}{N^2}\frac{N-1}{N+1} \int \big[\delta(h - c_n h_1 - s_n h_2) + \delta(h - c_n h_2 + s_n h_1)\big] q_H(h_1) q_H(h_2) \rd h_1 \rd h_2,
\end{multline}
and similarly for \(m\),
\begin{equation}
    -(m'-m) = \frac{2}{N+1} + 4 \frac{N-1}{N+1}\frac{N-m}{N} \left(\frac{N-m}{N} - 1 \right).
\end{equation}

In Eq.~\eqref{eqn:FullFlowq}, the nonlinear term in \(q_H\) is explicitly smaller by a factor of \(N\) than the other terms. We consider the large \(N\) limit and neglect this term, as well as other subleading terms in \(N\). We further cast the discrete update equation as a differential equation in \(n/N\), the number of Jacobi rotations per row, which has small increments of size \(1/N\). 
In this treatment, the cosines and sines, \(c_n\) and \(s_n\), become rapidly varying functions of \(n/N\), the fluctuations of which can be averaged over. Before this averaging, the differential form of the flow equation for \(q_H\) is
\begin{equation}\label{eqn:ApproxFlowq_NoAverage}
    \frac{\rd q_H(h)}{\rd(n/N)} = - 2 \delta(h-\mathrm{w}_n) - 4 q_H(h) + \frac{4}{c_n} q_H\left( \frac{h}{c_n}\right)  + \frac{2}{|s_n|}\left[q_H\left( \frac{h}{s_n}\right) + q_H\left(-\frac{h}{s_n}\right) \right],
\end{equation}
and performing the average gives
\begin{equation}\label{eqn:ApproxFlowq}
    \frac{\rd q_H(h)}{\rd(n/N)} = - 2 \delta(h-\mathrm{w}_n) - 4 q_H(h) + \int \frac{4}{c} q_H\left( \frac{h}{c}\right) P_c(c|n) \,\rd c + \int \frac{2}{|s|}\left[q_H\left( \frac{h}{s}\right) + q_H\left(-\frac{h}{s}\right) \right] P_s(s|n) \,\rd s.
\end{equation}
\end{widetext}
Here, \(P_c(c|n)\) is the distribution of the cosines \(c_m = \cos(\eta_m/2)\) over a window in the iteration number \(m \in [n-\Delta n, n+\Delta n]\), and similarly for \(P_s(s|n)\) and \(s_m\). The window size \(\Delta n\) should be small compared to \(N\), but still diverging. For instance, \(\Delta n = \sqrt{N}\).

Dropping the subleading term in \(N\) allows Eq.~\eqref{eqn:ApproxFlowq} to be solved analytically, as described below. 

While the sparse regime flow equation Eq.~\eqref{eqn:ApproxFlowq} appears linear, the maximum element \(w_n\) and angles \(\eta_n\) (via \(P_c\) and \(P_s\)) are implicitly determined by \(q_H\), so that the equation properly remains a nonlinear integrodifferential equation.
However, when considering a small window of Jacobi rotations in \(n\), the decimated elements \(w_n\) can be treated as monotonically decreasing~\cite{hahn2025predicting}, and the average effect of the rotation angles \(\eta_n\) can also be simply analyzed.
The maximum absolute element \(w\) can then parameterize the flow time (rather than \(n/N\)).

A convenient choice of variables is
\begin{align}\label{eqn:ChangeVarsq}
    \Gamma &= \log\frac{w_0}{w}, \quad
    \ell = \log\left|\frac{w}{h}\right|, \text{ and} \nonumber \\
    \tilde{q}_\ell(\ell| \Gamma) &= w_0 e^{-\ell-\Gamma} \left[ q_H(w_0 e^{-\ell-\Gamma}| n_\Gamma)\right.\\
    &\left.\qquad \qquad \quad+ q_H(-w_0 e^{-\ell-\Gamma}| n_\Gamma) \right]. \nonumber
\end{align}
Here, \(\Gamma\) parametrizes flow time (\(n_\Gamma\) is the corresponding index in the Jacobi algorithm) and \(\ell\) is a reparameterization of \(h \in [-w,w]\) to the fixed range \([0,\infty)\).
The definition of \(\tilde{q}_\ell\) follows from a change of variables from \((h,n)\) to \((\ell, \Gamma)\): the prefactor of \(h = w_0 e^{-\ell-\Gamma}\) comes from the Jacobian in the change of variables, and we add both \(q(h)\) and \(q(-h)\) because both positive and negative \(h\) map to the same \(\ell\).

To reexpress derivatives with respect to \(n\) in terms of the new variables, we use
\begin{equation}
    \frac{\rd \tilde{q}_\ell}{\rd n} = \frac{\rd \ell}{\rd n} \partial_\ell \tilde{q}_\ell + \frac{\rd \Gamma}{\rd n} \partial_\Gamma \tilde{q}_\ell
\end{equation}
and
\begin{equation}
    \frac{\rd \ell}{\rd n} = \frac{1}{w} \frac{\rd w}{\rd n},
    \quad
    \frac{\rd \Gamma}{\rd n} = -\frac{1}{w} \frac{\rd w}{\rd n} = \frac{2}{w \rhodec(w)}.
\end{equation}
The \(\rd n / \rd w\) derivatives can be expressed in terms of \(\rhodec\) using the assumption that \(w\) and \(n\) are monotonically related. In this case, Eq.~\eqref{eqn:ndec_integral} gives \(\rhodec = - 2\rd n/\rd w\). 

{Consider first Eq.~\eqref{eqn:ApproxFlowq_NoAverage}. Using the variables in Eq.~\eqref{eqn:ChangeVarsq} provides}
\begin{multline}\label{eqn:ApproxFlowqtilde}
    \partial_\Gamma \tilde{q}_\ell - \partial_\ell \tilde{q}_\ell + \delta(\ell) \tilde{q}_\ell(0) \\
    = 2 \tilde{q}_\ell(0)[\tilde{q}_\ell\left(\ell - \ell_c\right) + \tilde{q}_\ell\left(\ell - \ell_s\right) - \tilde{q}_\ell(\ell)],
\end{multline}
where \(\ell_c = \log(1/c)\) and \(\ell_s = \log(1/|s|)\). We have suppressed the dependence of \(\tilde{q}_\ell\) on \(\Gamma\).
The left-hand side of Eq.~\eqref{eqn:ApproxFlowqtilde} implements the rescaling of \(h\) by \(w\), which, in terms of the logarithmic variable \(\ell\), is a translation towards smaller values. The \(\delta\)-function on the left-hand side implements the decimation, ensuring that \(\tilde{q}_\ell(\ell < 0) = 0\) for all flow times \(\Gamma\), provided this is satisfied by the initial conditions.
Meanwhile, the right-hand side implements the Jacobi rotation, which in our current level of approximation simply removes elements at \(\ell\) and produces two new nonzero elements at \(\ell-\ell_c\) and \(\ell- \ell_s\).

Reintroducing the average over angles leads to the final form of the flow equation within the first level of approximation. Let 
\(P_c(\ell_c| \Gamma)\) and 
\(P_s(\ell_s| \Gamma)\) be the respective distributions of 
\(\ell_c\) and 
\(\ell_s\) at flow time \(\Gamma\). 
Performing the average over these distributions of \(\ell_c\) and \(\ell_s\) results in a convolution in Eq.~\eqref{eqn:ApproxFlowqtilde},
\begin{equation}\label{eqn:qlFlow}
    \partial_\Gamma \tilde{q}_\ell - \partial_\ell \tilde{q}_\ell + \delta(\ell) \tilde{q}_\ell(0) = 2 \tilde{q}_\ell(0)\left[(P_c+P_s)* \tilde{q}_\ell - \tilde{q}_\ell \right],
\end{equation}
where the convolution is in the \(\ell\) variable.

Equation~\eqref{eqn:qlFlow} can be solved exactly, with no further physically motivated assumptions or approximations.
The solution is presented in Appendix~\ref{app:FunctionalFlowSolutionAppendix}. (Indeed, the solution is subsumed by the more general solution to the perturbed flow equation of \autoref{sec:FunctionalFlowInstability}.)
Given the initial distribution of Hamiltonian matrix elements, \(\tilde{q}_\ell(\ell|0)\), the result for \(\tilde{q}_\ell(0|\Gamma) = w\rhodec/N\) is
\begin{equation}\label{eqn:gsol}
    \frac{w \rhodec(w)}{N} = \frac{\tilde{q}_\ell(\Gamma_w|0)}{1+ 2\int_0^{\Gamma_w} \rd \gamma\, \tilde{q}_\ell(\gamma|0)}.
\end{equation}

This solution does not express the resonance instability that we aim to capture. Indeed, consider an initial power-law distribution of matrix elements, which corresponds to an exponential \(\tilde{q}_\ell(\ell|0) = c e^{(1-\theta_0)\ell}\). We predict 
\begin{equation}
    \frac{\rhodec(w)}{N} = \frac{c e^{(1-\theta_0)\Gamma_w}/w}{1+2c  \frac{e^{(1-\theta_0)\Gamma_w} -1}{1-\theta_0}} = \frac{c (w/w_0)^{-2+\theta_0}/w_0}{1+2c  \frac{(w/w_0)^{-1+\theta_0} -1}{1-\theta_0}}.
\end{equation}
For \(\theta(w)\), we obtain
\begin{align}
    \theta(w) &= 1-\rd_{\Gamma_w} \log[\tilde{q}_\ell(0|\Gamma_w)]\nonumber \\
    &= \theta_0 + \frac{2 c (w/w_0)^{-1+\theta_0}}{1+2c\frac{(w/w_0)^{-1+\theta_0}-1}{1-\theta_0}}.
    \label{eq:theta_funct1}
\end{align}
Since $\theta_0 < 1$, the asymptotic value of \(\theta(w)\) for \(w \to 0\) is always \(\theta = 1\). Initially, negative values of \(\theta(w_0)\) fail to diverge, and there is no signature of resonance proliferation.

The lack of a resonance instability is because the first level of approximation is too aggressive, modeling matrix elements as always decreasing as soon as they become nonzero.
Indeed, the term in Eq.~\eqref{eqn:ApproxFlowq} encoding the Jacobi rotation [with \(\rho_H(h/c)/c\) and \(\rho_H(h/s)/s\)] takes an element \(h\) and replaces it with two elements \(c h\) and \(s h\). Both new elements are smaller than \(h\), and so are not decimated in the next Jacobi update.
The flow equation continually pushes rotated elements to smaller values of \(h\), such that they cannot have a large effect on the distribution of decimated largest elements.
As such, there is no influence of resonances---or any rotations---on \(\rhodec\), and thus no associated instability.

In the next section, we show that even a very small perturbation to the flow equation~\eqref{eqn:ApproxFlowq}, accounting for matrix elements sometimes increasing, allows resonances to proliferate.

\subsection{Second approximation: resonances can enhance couplings}
\label{sec:FunctionalFlowInstability}

In this section, we slightly perturb the flow equation from the first level of approximation, Eq.~\eqref{eqn:qlFlow}, and show that the solution changes drastically.
The perturbation to Eq.~\eqref{eqn:qlFlow} is motivated by the physical features lacking in Eq.~\eqref{eqn:qlFlow}---namely that resonances cannot enhance couplings between states in Eq.~\eqref{eqn:qlFlow}---but the particular way we model those features is motivated by maintaining analytic control. 
The result, Eq.~\eqref{eqn:qlFlowPerturbed}, recovers some of the qualitative features of the heuristic flow and the numerical results in \autoref{fig:theta_flow}. This includes the instability for $\theta<0$ and the line of fixed points $\theta(0)>0$. However, it misses other features, such as initial conditions $\theta(w_0)>0$ eventually flowing to the dense regime, and the quantitative BKT critical exponents predicted by the heuristic flow equation.

The flow equation Eq.~\eqref{eqn:qlFlow} does not allow matrix elements to increase along the flow (\(\ell\) is nondecreasing, aside from the trivial rescaling of coordinates). By contrast, in the actual implementation of the Jacobi algorithm---and in the full SJA functional flow equation, Eq.~\eqref{eqn:FullFlow}---individual matrix elements of \(H\) can occasionally increase. 
An extreme case occurs when a resonance couples two elements \(h_1,h_2\approx w\): for a perfect resonance with \(c=s=1/\sqrt{2}\), one finds \(h' = c h_1 + s h_2 \approx \sqrt{2}\,w\).
More generally, we have \(h'>h_1\) whenever \(h_2/h_1 > (1-c)/s \approx s/2\) (for \(s \ll 1\)), showing that even small rotation angles can lead to modest increases in matrix elements.

There are two theoretically well-motivated ways to incorporate processes that allow matrix elements to increase within the functional flow equation. The first would be to retain the subleading terms neglected in the flow equation~\eqref{eqn:FullFlowq}. However, we have not been able to analyze the resulting equations in a controlled manner. The second alternative would be to broaden the \(\delta\)-function appearing in the decomposition Eq.~\eqref{eqn:SparseFormRho}, while still neglecting subleading in $N$ contributions in Eq.~\eqref{eqn:FullFlowq}. This would permit matrix elements to grow through the rotation of a large element \(h_1\) with a small one \(h_2\), while still excluding the rotation of two comparably large elements. Although we expect such an approximation to be adequate in the sparse regime, it likewise resists controlled analysis.

Instead, we adopt a minimal phenomenological modification that preserves analytic tractability. Specifically, we replace the strict decimation of matrix elements at \(\ell=0\) by a distribution of decimations over all \(\ell\).
The interpretation of this modification is as follows. Due to some processes not accounted for in the flow equation, matrix elements can sometimes increase such that they eventually reach \(|h|=w\) (\(\ell=0\)), and are decimated. 
As our flow equation ignores those processes, the elements that should actually be decimated are still inside the bulk of the distribution \(|h| < w\) (\(\ell > 0\)) within our treatment.
The distribution of decimations over all values of \(\ell\) accounts for this by decimating matrix elements in the bulk of the distribution, which should have been able to reach \(\ell=0\).
Accordingly, for the purpose of calculating \(\rhodec\) and the rotation angles \(\eta\), these bulk elements are treated as having reached \(\ell=0\).

We assume a number distribution of decimated elements
\begin{equation}\label{eqn:ndecAssumption}
    n^\epsilon_{\mathrm{dec}}(\ell| \Gamma) = \frac{e^{-\ell/\epsilon}}{\epsilon} \tilde{q}_\ell(\ell; \Gamma)
\end{equation}
where \(\epsilon > 0\) is a small parameter and \(\tilde{q}_\ell\) now has support for all \(\ell\).
Without supposing the functional form of \(n^\epsilon_{\mathrm{dec}}(\ell| \Gamma)\), the scale \(w = |\mathrm{w}|\) is now defined by the condition \(n^\epsilon_{\mathrm{dec}}(h=\mathrm{w}| \Gamma) = q_H(h=\mathrm{w}| \Gamma)/\epsilon\).
The norm of \(n_{\mathrm{dec}}\) is defined to be
\begin{equation}\label{eq:int_ndec}
    \int_{-\infty}^{\infty}n_{\mathrm{dec}}(\ell| \Gamma) \rd \ell =: Q(\epsilon|\Gamma).
\end{equation}

Additionally, we make a small-angle approximation, assuming
\begin{equation}\label{eq:small_angle}
    \cos\frac{\eta}{2} \approx 1.
\end{equation}
As the small-angle approximation only improves at large \(\Gamma\) (typical rotation angles scale as \(e^{-\Gamma}\)), this approximation should be asymptotically well-controlled. At the level of Eq.~\eqref{eqn:qlFlow}, this approximation amounts to \(P_c(\ell) = \delta(\ell)\).

Incorporating Eq.~\eqref{eq:int_ndec} and~\eqref{eq:small_angle}, the functional flow equation~\eqref{eqn:qlFlow} takes the form
\begin{equation}\label{eqn:qlFlowPerturbed}
    \partial_\Gamma \tilde{q}_\ell 
    - \partial_\ell \tilde{q}_\ell + n_{\mathrm{dec}} = 2 Q(\epsilon) P_s * \tilde{q}_\ell.
\end{equation}
If \(\tilde{q}_\ell(\ell)\) is supported on \(\ell \geq 0\), then taking \(\epsilon \to 0\) recovers Eq.~\eqref{eqn:qlFlow}, up to the small angle approximation.

Although the specific perturbation in Eq.~\eqref{eqn:ndecAssumption} is not quantitatively accurate, it probes the stability or instability of the solution in Eq.~\eqref{eq:theta_funct1}.
If the solutions to Eq.~\eqref{eqn:qlFlowPerturbed} are very different from those of Eq.~\eqref{eqn:qlFlow}, it indicates that Eq.~\eqref{eqn:qlFlow} had an instability, and we should not trust its solution.

In Appendix~\ref{app:FunctionalFlowSolutionAppendix}, we find an approximate solution to Eq.~\eqref{eqn:qlFlow} which is valid when \(\epsilon\) is small and \(\theta\) is bounded.
The requirement that \(\theta\) be bounded means that the solution is not guaranteed to be accurate when \(\theta\) is diverging, but we believe that such a divergence should nonetheless signify delocalization.
The solution is given by
\begin{equation}\label{eqn:PerturbedSolutionMain}
    \frac{w \rhodec(w)}{N} \approx \frac{\tilde{q}_\ell(\Gamma_w|0)}{1- A_\epsilon \int_0^{\Gamma_w} \rd \gamma\, e^{-\gamma} \tilde{q}_\ell(\gamma|0)},
\end{equation}
where \(A_\epsilon>0\) is a constant (independent of \(w\)) which is exponentially small in \(\epsilon\). 
This solution diverges for some \(\Gamma\) whenever
\begin{equation}\label{eqn:DivergenceCondition}
    \int_0^\infty \rd \gamma\, e^{-\gamma} \tilde{q}_\ell(\gamma|0) > A_\epsilon^{-1}.
\end{equation}
If \(\tilde{q}_\ell(\ell|0) = c e^{(1-\theta_0)\ell}\), then the solution diverges when \(\theta_0 < 0\) for any value of $c$. On the other hand, if \(\theta_0>0\), there is always a small enough \(c\) such that the solution remains finite.
That is, any negative value of \(\theta\) has an instability towards resonance proliferation, while positive values of \(\theta\) can be stable.

Indeed, for \(\tilde{q}_\ell(\ell|0) = c e^{(1-\theta_0)\ell}\), \(\theta(w)\) eventually diverges when
\begin{equation}\label{eqn:theta0_critical}
    \theta_0 \leq c A_\epsilon,
\end{equation}
so that even small positive values of \(\theta_0\) can belong to the thermalizing phase.
Note that \(\theta_0\) is defined by the initial distribution \(\tilde{q}_\ell(\ell|0)\), and not by \(\theta(w_0)\), set by the log-slope of \(\rhodec\). In this case, \(\theta_0 \neq \theta(w_0)\).

The full forms of \(\rhodec\) and \(\theta(w)\) are given by
\begin{equation}\label{eqn:PerturbedFunctionalrhodec}
    \frac{\rhodec(w)}{N} = \frac{ c (w/w_0)^{-2+\theta_0}/w_0 }{1 + c A_\epsilon \frac{(w/w_0)^{\theta_0}-1}{\theta_0} },
\end{equation}
and
\begin{equation}
    \label{eq:solution_functional}
    \theta(w) = \theta_0 - \frac{ c A_\epsilon (w/w_0)^{\theta_0} }{1 + c A_\epsilon \frac{(w/w_0)^{\theta_0}-1}{\theta_0} }.
\end{equation}%
\begin{figure}
    \centering
    \includegraphics[width=\linewidth]{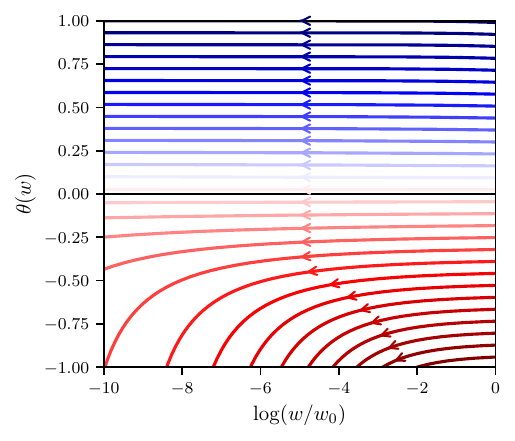}
    \caption{Plot of the solution of the functional flow equations, reported in Eq.~\eqref{eq:solution_functional}. \emph{Parameters}: $cA_{\epsilon} = 0.01$.}
    \label{fig:sparse_theta}
\end{figure}%
The plot of the solution is shown in~\autoref{fig:sparse_theta}. This solution expresses the qualitative features we identified in the heuristic flow equations. Any \(\theta \in [0,1]\) can occur as an asymptotic value of the flow, while any flow towards \(\theta < 0\) is interrupted by a divergence to \(\theta \to -\infty\) at a finite value \(w = w_c\), 
\begin{equation}\label{eqn:FunctionalCriticalwc}
    \frac{w_c}{w_0} = \left(1-\frac{\theta_0}{c A_\epsilon} \right)^{1/\theta_0}.
\end{equation}

However, there are other features of the heuristic flow equations which are not reproduced by Eq.~\eqref{eq:solution_functional}.
While Eq.~\eqref{eq:solution_functional} exhibits a line of fixed points, it does not reproduce the BKT critical exponents of the heuristic flow equation~\eqref{eq:Heuristic}. From Eq.~\eqref{eqn:FunctionalCriticalwc}, \(w_c/w_0\) approaches zero as a power law as \(c A_\epsilon \to \theta_0\), where the power depends on \(\theta_0\). This is in contrast to the BKT scaling, which gives non-power-law behavior of the critical scaling.
Based on other analyses of Hilbert space localization, for instance in the RRG model~\cite{vanoni2023renormalization}, it seems unlikely that the true critical behavior is actually a power law.

The solution does not allow for initially positive values of \(\theta(w_0)\) to flow to negative values of \(\theta(w)\).
[Note that this is not in tension with the critical point for \(\theta_0\) in Eq.~\eqref{eqn:theta0_critical} being strictly positive, as positive \(\theta_0 >0\) can still result in \(\theta(w_0) \leq 0\).]
Further, taking \(\theta_0 > c A_\epsilon\) in Eq.~\eqref{eq:solution_functional} produces a \(\theta(w)\) which increases as \(w\) decreases.
Neither of these properties is reproduced by the heuristic flow equations or our numerics in \autoref{sec:numerics}, indicating that the functional flow equation Eq.~\eqref{eqn:qlFlowPerturbed}, while more accurate than Eq.~\eqref{eqn:qlFlow}, continues to miss important features.
It is conceivable that a different form of the perturbation, closer to that motivated by the subleading terms in Eq.~\eqref{eqn:FullFlowq}, may correct these problems.

Nonetheless, a broad lesson of this analysis is that a process where matrix elements increase under Jacobi rotations is necessary to observe the resonance proliferation instability.
That is, the fact that a resonance between two states can result in dressed states with enhanced couplings (matrix elements) to some third state is important to the process of thermalization.

\section{Observable predictions}
\label{sec:ObservablePredictions}

The SJA predicts qualitative features of the phase diagram of the LRP, RRG, and the toy model of locally perturbed MBL: The existence of both a stable localized and delocalized phase, separated by a sharp transition.

\begin{figure}
    \centering
    \includegraphics[width=\linewidth]{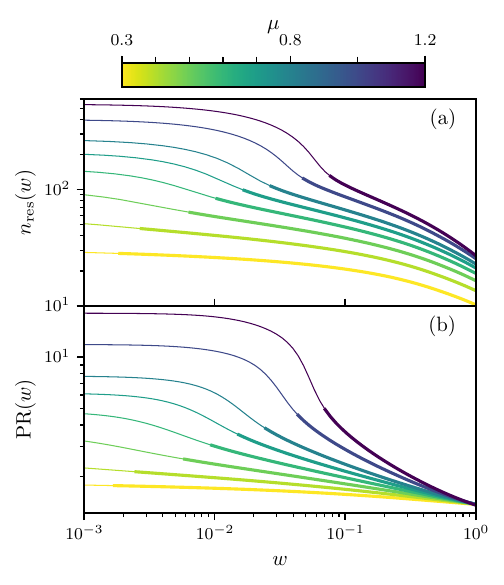}
    \caption{(a) Number of resonances $n_{\mathrm{res}}(w)$ as a function of $w$, obtained numerically for the LRP with $N=1024$ and different values of $\mu$ as indicated by the tick marks on the colorbar.
    (b) Average participation ratio of the states $\ket{i_{n(w)}}$ for the same model and parameters, showing good agreement with the behavior of $n_{\mathrm{res}}(w)$, in particular in the position of the “knee" (where the thick lines end).}
    \label{fig:nresvsPR}
\end{figure}

\begin{figure}
    \centering
    \includegraphics[width=\linewidth]{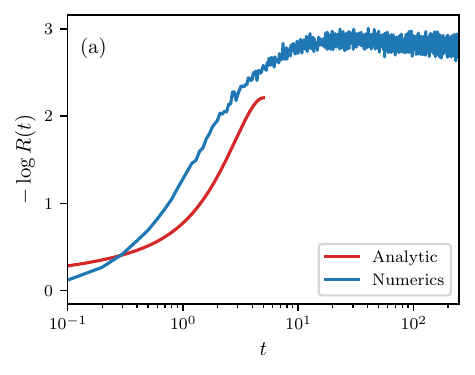}
    \includegraphics[width=\linewidth]{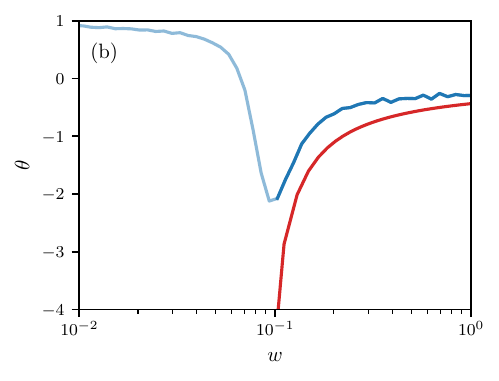}
    \caption{(a) The analytical prediction for the return probability obtained from the solution of the functional flow equations (red) compares well to the numerical results (blue) obtained from exact time evolution of the Lévy-RP model, with $\mu = 1.2$. \emph{Parameters}: $c=0.013$, $A_{\epsilon}=9$ and $\theta_0 = 1-\mu = -0.2$. (b) Comparison between the prediction for $\theta$ given by the solution of the functional flow equation (red) and the numerical result (blue), with the same parameters as in panel (a).}
    \label{fig:funct_retprob}
\end{figure}

In this section, we relate the predictions of the SJA to more quantitative physical probes of localization and thermalization by expressing observables such as the return probability and time-dependent correlation functions in terms of the Jacobi flow variables (see~\autoref{fig:nresvsPR}, Eq.~\eqref{eq:Ct_scaling}, and~\autoref{fig:funct_retprob}).

The number of resonances per state above scale \(w\), \(n_{\mathrm{res}}(w)/N\) [Eq.~\eqref{eqn:nres_def}], is a proxy for the participation ratio (PR, defined below), and is thus a diagnostic for the delocalization of states in Hilbert space (\autoref{fig:fig3}).
Each time a state encounters a resonance during the Jacobi flow, its probability mass is split roughly evenly between the two basis states involved in the rotation.
The average participation ratio (PR) is defined as
\begin{equation}
\label{eq:ret_prob_def}
    \mathrm{PR}(w) = \left( \frac{1}{N}\sum_{i,j=1}^N |\braket{i_{0}}{j_{n(w)}}|^4\right)^{-1},
\end{equation}
and is equal to $N_s$ for a uniformly delocalized state over $N_s$ $n=0$ basis states. 
We thus expect that $\mathrm{PR}(w) \propto n_{\mathrm{res}}(w)/N$, at least early on in the Jacobi algorithm in the sparse regime, when the probability of a resonance is small, and the probability that the resonance involves already resonant states in the $n=0$ basis is even smaller.

\autoref{fig:nresvsPR} shows that \(n_{\mathrm{res}}(w)\) and $\mathrm{PR}(w)$ have the same shape in the LRP model. In particular, the “knee" signaling the onset of the dense regime occurs at the same value of $w$ in both quantities.

Beyond eigenstate properties, \(n_\mathrm{res}(w)/N\) also predicts certain aspects of real-time dynamics.
In the regime where \(\theta(w)\) is slowly varying, well before it diverges and we cross over into the dense regime, we predict stretched exponential decay with a slowly varying exponent,
\begin{equation}
\label{eq:Ct_scaling}
    C(t) \propto e^{-(t/\tau)^{\beta(t)}}.
\end{equation}
Our argument builds on that in Ref.~\cite{Long2023Phenomenology}.

Ref.~\cite{Long2023Phenomenology} found that infinite-temperature autocorrelation functions \(C(t)\) of observables that are diagonal in the disorder basis can be approximated as
\begin{equation}\label{eqn:rhoresToCt}
    C(t) \propto \exp\left[ \frac{1}{N}\int_{w_c}^{w_0} \rhores(w) \cos(w t) \,\rd w \right],
\end{equation}
in the regime \(w_0^{-1} \ll t \ll w_c^{-1}\), where \(w_c\) is a lower cutoff on the decimated matrix elements, set by the crossover to the dense regime.
That is, the log of the autocorrelator is, up to a constant, the Fourier transform of the distribution of resonances.
[Eq.~\eqref{eqn:rhoresToCt} only accounts for the effect of resonances. There should be quantitative corrections due to smaller Jacobi rotations.]
From Eq.~\eqref{eqn:rhoresToCt} and the ansatz \(\rhores(w) \propto w^{-1+\theta}\), it was concluded in Ref.~\cite{Long2023Phenomenology} that the autocorrelation functions in prethermal MBL decay as a stretched exponential \(e^{-(t/\tau)^{\beta}}\) with an exponent \(\beta = -\theta\). The Jacobi exponent is not intended to replace the usual RRG observables. Instead, it provides a scale-resolved measure of resonance formation whose integrated behavior can be compared with more conventional quantities.

Consider the ansatz that accounts for resonance formation at previous scales, \(\rhores(w) \propto w^{-1+\theta(w)}\). Alternatively, we have the prediction of the functional flow equations, Eq.~\eqref{eqn:PerturbedFunctionalrhodec}. 
In the regime where \(\theta(w)\) is slowly varying, well before it diverges and the algorithm crosses over into the dense regime, we thus predict stretched exponential decay with a slowly varying exponent, Eq.~\eqref{eq:Ct_scaling}.
Near the divergence of \(\theta(w)\), \(\rhores(w)\) no longer resembles a power law over a large range of scales \(w\), and its Fourier transform is similarly not well described as a power law with a flowing exponent. 
Instead, some regularization from the crossover to the dense regime takes over, and the asymptotic decay of autocorrelators and return probabilities presumably becomes exponential.
On the localized side, where \(\theta >0\), Eq.~\eqref{eqn:rhoresToCt} predicts a power-law decay of autocorrelators to a constant, which is the expected behavior.

A similar formula to Eq.~\eqref{eqn:rhoresToCt} applies to return probabilities,
\begin{equation}
    R(t) = |\braket{\psi_0}{\psi(t)}|^2.
\end{equation}
The prediction of SJA is also that \(R(t)\) decays as a stretched exponential with a slowly varying exponent. Indeed, stretched exponential decay has also been observed in the return probability in the RRG~\cite{bera2018return}.

In \autoref{fig:funct_retprob}, we compare the combination of the functional flow equation prediction for \(\rhores(w)\) from Eq.~\eqref{eqn:PerturbedFunctionalrhodec} and the autocorrelator expression in Eq.~\eqref{eqn:rhoresToCt} to the numerically calculated return probability in the LRP model, finding good agreement.
We treat \(w_0\), \(c\) and \(A_\epsilon\) in Eq.~\eqref{eqn:PerturbedFunctionalrhodec} as fit parameters, and take \(\theta_0 = 1-\mu\) from the LRP model parameters. 

We now turn to predictions regarding the transition between localized and delocalized phases.
The heuristic (\autoref{sec:HeuristicFlow}) and functional (\autoref{sec:FunctionalFlow}) flow equations make distinct predictions for critical behavior, so that a reliable conclusion regarding critical behavior cannot be made based on our analysis. 
Given that the functional flow equation predicts more familiar power-law critical behavior, and that this prediction is disfavored by independent analysis~\cite{vanoni2023renormalization}, we focus on the BKT scaling prediction of the heuristic flow equation.
This scaling predicts that the total number of resonances \(n_{\mathrm{res}}(w=0)/N\), and hence the average PR of eigenstates, diverges at the transition faster than any power law, in agreement with predictions for the divergence of the localization length in the RRG~\cite{vanoni2023renormalization,tikhonov2021AndersonMBL}.

Finally, we address the thermalization time, \(\tau\). On the thermal side, we interpret \(w_c\), the value of \(w\) at which \(\theta(w)\) diverges, as an inverse thermalization time.
Strictly speaking, to convert \(w_c\) to \(\tau\), one should use Eq.~\eqref{eqn:betaw_def}. Comparing \(w_c\) to the inverse thermalization time is coarse dimensional analysis, which we nonetheless expect to capture the relevant scaling at fixed system size.
In the heuristic flow equations, the divergence of \(w_c^{-1}\) as \(w_0\) and \(\theta_0\) are varied across the separatrix reflects the BKT scaling, giving a thermalization time that diverges faster than any power law. Specifically, the analysis of Ref.~\cite{1973JPhC....6.1181K} predicts that, when \(\theta_0\) crosses its critical value \(\theta_c\) from below (the thermal side) at a fixed \(w_0\), the critical scale \(w_c\) goes to zero as
\begin{equation}
    \log (-\log w_c) \sim \sqrt{- \log (\theta_c- \theta_0)}.
\end{equation}
(All logarithms are base \(e\).)
Assuming that the thermalization time \(\tau\) scales as \(\log \tau \sim -\log w_c\), and that the dependence of \(\theta_0\) on the microscopic disorder strength \(W\) can be linearized near the separatrix, \(\theta_c - \theta_0 \sim (W_c - W)/A\) (where \(W_c\) is the critical value of \(W\) and \(A\) is a positive constant), we find
\begin{equation}
    \log \log\tau \sim \sqrt{\log \frac{A}{W_c-W}}.
\end{equation}
That is, \(\log \tau\) diverges as \(\exp\{[\log (A/\Delta W)]^{1/2}\}\) (up to subleading factors) where \(\Delta W = W_c - W\). The divergence of \(\log \tau\) is slower than any power law. However, \(\tau\) itself diverges roughly as \(\exp\exp\{[\log (A/\Delta W)]^{1/2}\}\), which is faster than any power law.

\section{Discussion}
\label{sec:discussion}

Numerically extracted critical properties in localization transitions often change significantly with system size.
This size dependence highlights the need for a well-founded notion of renormalization flow for quantum dynamics.
The Jacobi algorithm and statistical Jacobi approximation (SJA), while not strictly renormalization procedures as they do not remove degrees of freedom, aim to meet this need for a scale-dependent treatment of quantum dynamics.

We have demonstrated the application of the SJA to models exhibiting Hilbert-space or real-space localization: the L\'evy-Rosenzweig-Porter (LRP) random matrix model, the Anderson model on a regular random graph (RRG), and the standard disordered Heisenberg model for many-body localization (MBL).
The same Jacobi diagnostic can be applied to all three systems, although its physical interpretation differs between the Hilbert-space localization problems and the real-space XXZ chain. It predicts the flow of an exponent \(\theta(w)\) controlling resonances. When resonances proliferate, \(\theta\) flows to large and negative values as smaller Rabi frequencies are resolved. When not, \(\theta\) flows to a stable positive value, so that the localized phase is characterized by a line of fixed points with $0<\theta\leq 1$. The separatrix that flows to $\theta(0)=0$ defines the transition. These qualitative predictions are borne out by numerical application of the Jacobi algorithm in all three models. We comment more on the application to MBL below.

Quantitative aspects of the flow, and the nature of the phase transition depend on the approximations used to solve the SJA functional flow equation, Eq.~\eqref{eqn:FullFlow}. We presented two approximations in \autoref{sec:HeuristicFlow} and \autoref{sec:FunctionalFlow}. The first, more heuristic approach produces BKT-like scaling, whereas the functional approximation produces power-law scaling. Their shared prediction is the instability toward resonance proliferation; the critical scaling remains unresolved.

The SJA also predicts the behavior of more traditional probes of eigenstate localization, such as eigenstate participation ratios, and dynamical signatures of thermalization, such as the functional form of return probabilities.

Many-body localized systems are not Hilbert-space localized. Their asymptotic instability to delocalization is also believed to be caused by thermal avalanches, not by many-body resonances, at least with random disorder. Nevertheless, the SJA can describe dynamics in various regimes.

Consider the regime around $W\approx 7$ in the random-field Heisenberg chain. At the system sizes accessible here, standard diagnostics and the initial Jacobi flow display localized-like behavior, while the later flow shows increasing resonance formation. We refer to this finite-size and finite-time window as prethermal as a working interpretation. The SJA flows in which $\theta(w_0)$ starts from a bare positive value (so that the system initially appears localized and stable to resonance formation) and then flows, with decreasing $w$, to negative values of $\theta$ (so that resonances have proliferated and the system is delocalized) provide a scale-resolved description of this finite-size regime, as we have shown in Fig.~\ref{fig:XXZ_W7}.

Next, consider the system at much larger disorder strengths (say $W \approx 15$ in the well-studied disordered Heisenberg chain). Simple estimates~\cite{Crowley2022Constructive} place this regime in the localized phase of the SJA flow, so that despite the renormalization of $\theta$, typical regions of the system are localized, and local perturbations induce, at best, the formation of a few short-range resonances in the unperturbed eigenbasis. However, the thermodynamic system is believed to be thermalizing at these disorder strengths~\cite{Morningstar2022Avalanches, Sels2022avalanches}. The instability leading to delocalization here is avalanches. A single framework that accounts for resonance physics in typical regions and avalanches induced by rare regions is outstanding; such a framework would also predict critical properties of the MBL transition in the thermodynamic limit.

The question of whether the SJA can describe MBL with correlated disorder, including quasiperiodic disorder~\cite{Iyer2013quasiperiodic,Khemani2017twoclasses,Agrawal2020quasiperiodic,Agrawal2022dgt1,Tu2023qpavalanche,Tu2024quasiperiodicprethermal}, is still open.
With correlated disorder, rare low-disorder regions may never occur, so that avalanches may not destabilize typically localized regions.
In this case, it is conceivable that the proliferation of many-body resonances represents the leading instability~\cite{Znidaric2018Quasiperiodic}, and that the SJA analysis, or some modification thereof, can correctly describe the transition.
We leave this question for future work.

Our analysis can be extended to incorporate mobility edges in the spectrum of the studied models. 
In our numerics and analytical treatment, we assume infinite temperature by weighting all states equally. For instance, \(\rhodec(w)\) is a histogram of all decimated matrix elements, regardless of their energy.
To correctly find the mobility edge present in, for instance, the Anderson model on the RRG~\cite{Biroli2010bethe}, the energy of the states affected by Jacobi should also be included in the analysis.
In principle, the energy of the decimated states \(\ket{a_n}\) and \(\ket{b_n}\) can also be extracted from the Jacobi algorithm, and assembled into an energy-resolved distribution of decimated elements \(\rhodec(w, E_a, E_b)\)~\cite{Long2023Fermi}, which could be used to identify mobility edges.

\acknowledgments%
{%
We are grateful to Antonello Scardicchio for collaboration at the initial stages of this work, and to David A.\ Huse and Gil Refael for many useful discussions.
CV is grateful to Boston University for the kind hospitality during the initial stages of this work.
Data presented in this work, and the code used to generate it, are available on GitHub~\cite{GitHub}.
The work was supported in part by NSF DMR-1752759. DML is supported by a Stanford Q-FARM Bloch Fellowship, the US Department of Energy, Office of Science (Award No. DE-SC0019380), and a Packard Fellowship in Science and Engineering (PI: Vedika Khemani).
}

\appendix
\section{Solution to the functional flow equations}
\label{app:FunctionalFlowSolutionAppendix}

In \autoref{sec:FunctionalFlow}, we discussed sparse regime flow equations for the evolution of the distribution of Hamiltonian matrix elements throughout the Jacobi flow.
In this appendix, we present the detailed derivation of these flow equations.

Both the flow equation obtained in \autoref{sec:FunctionalFlowConstruction} and the perturbation thereto formulated in \autoref{sec:FunctionalFlowInstability} are of the general form
\begin{equation}\label{eqn:qlFlowAppendix}
    \partial_\Gamma \tilde{q}_\ell - \partial_\ell \tilde{q}_\ell + n^\epsilon_{\mathrm{dec}} \tilde{q}_\ell =  Q(\epsilon) P * \tilde{q}_\ell,
\end{equation}
where \(\tilde{q}_\ell(\ell|\Gamma)\) is the number density of the logarithmic matrix elements \(\ell = \log|w/h|\) per row, \(\Gamma = \log w_0/w\) acts as a flow time,
\begin{equation}
    n^\epsilon_{\mathrm{dec}}(\ell| \Gamma) = \frac{e^{-\ell/\epsilon}}{\epsilon} \tilde{q}_\ell(\ell| \Gamma)
\end{equation}
is the distribution according to which elements are decimated, and \(Q(\epsilon)\) is the norm of \(n^\epsilon_{\mathrm{dec}}(\ell| \Gamma)\). For \(\epsilon \to 0\) we have \( n^\epsilon_{\mathrm{dec}}(\ell) = \delta(\ell) \tilde{q}_\ell(0)\).
The term \(P * \tilde{q}_\ell\) is a convolution of \(P\) with \(\tilde{q}_\ell\), and encodes the effect of the Jacobi rotations. In \autoref{sec:FunctionalFlowConstruction} we take \(P(\ell | \Gamma) = 2[P_c(\ell)+P_s(\ell) - \delta(\ell)]\) (where \(P_c\) and \(P_s\) are the probability distributions of \(\ell_c = -\log\cos\tfrac{\eta}{2}\) and \(\ell_s = -\log\sin\tfrac{\eta}{2}\) respectively, which depend on \(\Gamma\)), while in \autoref{sec:FunctionalFlowInstability} we make a small angle approximation and use just \(P = 2 P_s\). For now, we will leave \(P\) unspecified.
The resulting solution of the flow equation will be exact for \(\epsilon =0\), and correct to leading order for \(0 < \epsilon \ll 1\).

Equation~\eqref{eqn:qlFlowAppendix} resembles the flow equations obtained in strong-disorder real-space RG~\cite{Fisher1992Random,fisher1994random}, and we can try to apply the same methods of solution.
Namely, we perform a two-sided Laplace transform. It is also convenient to use inverse variables for the transform (temperature \(T\) rather than inverse temperature \(\beta\)), that is, we define
\begin{equation}
    Q(T| \Gamma) = \frac{1}{T} \int_{-\infty}^\infty e^{-\ell/T} \tilde{q}_\ell(\ell| \Gamma) \,\rd \ell,
\end{equation}
(this agrees with the \(Q(\epsilon)\) we already defined) and similarly
\begin{equation}
    f(T| \Gamma) = \int_{-\infty}^\infty e^{-\ell/T} P(\ell| \Gamma) \,\rd \ell_c.
\end{equation}
While we have formally defined the integrals here as being over all real values of \(\ell\), making these genuine two-sided Laplace transforms, all our distributions of interest in \autoref{sec:FunctionalFlowConstruction} are supported only on non-negative values of \(\ell\), so these expressions reduce to the usual one-sided Laplace transform.

We use the following properties of the Laplace transform:
\begin{itemize}
    \item Convolution:
    \begin{equation}
        \frac{1}{T} \int_{-\infty}^\infty e^{-\ell/T} [P*\tilde{q}_\ell](\ell| \Gamma) \,\rd \ell = f(T|\Gamma) Q(T|\Gamma).
    \end{equation}

    \item Derivative (integrating by parts):\footnote{The one-sided Laplace transform of a derivative involves the initial value \(\tilde{q}_\ell(0)\). In our two-sided formulation, this term arises from the \(\delta\)-function in Eq.~\eqref{eqn:qlFlowAppendix} with \(\epsilon=0\).}
    \begin{equation}
    \label{eq:derivativeLT}
        \frac{1}{T} \int_{-\infty}^\infty e^{-\ell/T} \partial_l \tilde{q}_\ell(\ell| \Gamma) \,\rd \ell = \frac{1}{T}Q(T|\Gamma).
    \end{equation}

    \item Low temperature limit:
    \begin{equation}
        \lim_{T\to 0}Q(T) = \tilde{q}_\ell(\ell=0).
    \end{equation}
\end{itemize}
Applying the Laplace transform to Eq.~\eqref{eqn:qlFlowAppendix}, we obtain
\begin{multline}
    \partial_\Gamma Q(T) = \frac{1}{T}\left[Q(T) - \left(1+\tfrac{\epsilon}{T}\right)^{-1}Q\left(\tfrac{T\epsilon}{T+\epsilon}\right)\right] \\
    + f(T) Q(\epsilon) Q(T).
\end{multline}
We are interested in the limit of small \(\epsilon/T\), so we take the leading order expression
\begin{equation}\label{eqn:QTAppendix}
    \partial_\Gamma Q(T) = \frac{1}{T}\left[Q(T) - Q(\epsilon)\right]
    + f(T) Q(\epsilon) Q(T).
\end{equation}

If we treat \(Q(\epsilon| \Gamma)\) as known, Eq.~\eqref{eqn:QTAppendix} can be solved exactly for \(T > \epsilon\). We attempt a self-consistent solution, where we assume a functional form of \(Q(\epsilon| \Gamma) =: g(\Gamma)\), solve for \(Q(T| \Gamma)\), and then demand that \(\lim_{T \to \epsilon} Q(T| \Gamma) = g(\Gamma)\). This self-consistency condition can be solved in the limit of small \(\epsilon\).

Equation~\eqref{eqn:QTAppendix} is an affine ordinary differential equation (ODE) for \(Q(T)\), for which there is a standard method of solution. First consider just the part of Eq.~\eqref{eqn:QTAppendix} which is linear in \(Q(T)\),
\begin{equation}
    \partial_\Gamma Q(T| \Gamma) = \left[\frac{1}{T} + f(T| \Gamma) g(\Gamma)\right] Q(T| \Gamma),
\end{equation}
whose solution is
\begin{equation}
    Q(T| \Gamma) = \exp\left[ \frac{\Gamma}{T} + F(T| \Gamma) \right] Q(T| 0),
\end{equation}
where
\begin{equation}
    F(T| \Gamma) = \int_0^\Gamma \rd \gamma\, f(T| \gamma) g(\gamma).
\end{equation}
To solve the full Eq.~\eqref{eqn:QTAppendix}, we consider the ``comoving'' solution with respect to the linear part,
\begin{equation}
    X(T| \Gamma) := \exp\left[ -\frac{\Gamma}{T} - F(T| \Gamma) \right] Q(T| \Gamma).
\end{equation}
This satisfies the equation
\begin{align}
    \partial_\Gamma X(T| \Gamma) &= \left[ -\frac{1}{T} - f g \right] X + \left( \left[\frac{1}{T} + f g \right] X  - \frac{g}{T}e^{-\tfrac{\Gamma}{T} - F}\right) \nonumber\\
    &= - \frac{g}{T}e^{-\tfrac{\Gamma}{T} - F},
\end{align}
where the term in parentheses comes from differentiating \(Q\) in the product rule for \(X\), and the first term comes from differentiating the exponential.
This equation can be directly integrated,
\begin{equation}
    X(T| \Gamma) = X(T| 0) - \int_0^\Gamma \rd \gamma \frac{g(\gamma)}{T}\exp\left[ -\frac{\gamma}{T} - F(T| \gamma) \right].
\end{equation}
Reexpressed in terms of \(Q(T| \Gamma)\), we have
\begin{multline}\label{eqn:QTsol_Q0known}
    Q(T| \Gamma) = \exp\left[\frac{\Gamma}{T} + F(T| \Gamma) \right] Q(T| 0) \\
    - \int_0^\Gamma \rd \gamma \frac{g(\gamma)}{T}\exp\left[\frac{\Gamma - \gamma}{T} + F(T| \Gamma) - F(T| \gamma) \right].
\end{multline}
A slightly more convenient form is obtained by recalling that
\begin{equation}
    Q(T| 0) = \frac{1}{T}\int_0^\infty \rd \gamma\, e^{-\gamma/T} \tilde{q}_\ell(\gamma| 0),
\end{equation}
where we used the integration variable \(\gamma\) rather than \(\ell\). Substituting this, Eq.~\eqref{eqn:QTsol_Q0known} becomes
\begin{multline}\label{eqn:QTsubsLaplace}
    Q(T| \Gamma) = \frac{e^{F(T|\Gamma)}}{T}\int_0^\infty \rd \gamma\, e^{(\Gamma-\gamma)/T}\big[ \tilde{q}_\ell(\gamma|0) \\
    - e^{-F(T|\gamma)} g(\gamma) 1_{[0,\Gamma]}(\gamma) \big].
\end{multline}

As explained above, we now wish to take a \(T \to \epsilon\) limit at fixed \(\Gamma\), and demand that \(Q(\epsilon|\Gamma) = g(\Gamma)\). 
This gives an integral equation for \(Q(\epsilon|\Gamma)\),
\begin{multline}\label{eqn:QTsubsLaplaceEps}
    Q(\epsilon; \Gamma) = \frac{e^{F(\epsilon|\Gamma)}}{\epsilon}\int_0^\infty \rd \gamma\, e^{(\Gamma-\gamma)/\epsilon}\big[ \tilde{q}_\ell(\gamma|0) \\
    - e^{-F(\epsilon|\gamma)} Q(\epsilon| \gamma) 1_{[0,\Gamma]}(\gamma) \big].
\end{multline}
We cannot solve this in general, but a controlled approximation to the solution is given by
\begin{equation}
    Q(\epsilon|\gamma) = e^{F(\epsilon|\gamma)} \tilde{q}_\ell(\ell = \gamma| \Gamma = 0).
\end{equation}
Indeed, the relative error in Eq.~\eqref{eqn:QTsubsLaplaceEps} from this substitution is
\begin{equation}
    \frac{1}{\epsilon}\int_\Gamma^\infty \rd \gamma\, e^{(\Gamma-\gamma)/\epsilon} \left[\frac{\tilde{q}_\ell(\gamma|0)}{\tilde{q}_\ell(\Gamma|0)} -1 \right]
    = \epsilon \partial_\Gamma \log \tilde{q}_\ell(\Gamma|0) + O(\epsilon^2),
\end{equation}
so this approximation improves as \(\epsilon\) becomes small, provided \(\tilde{q}_\ell\) grows at most exponentially. That is, provided that \(\theta\) is finite.
The solution is exact for \(\epsilon = 0\).

We are left with the non-linear integral equation for \(g(\Gamma) = Q(\epsilon|\Gamma)\),
\begin{equation}
    g(\Gamma) e^{-\int_0^\Gamma \rd \gamma f(\epsilon|\gamma) g(\gamma)} =  \tilde{q}_\ell(\Gamma|0).
\end{equation}
This can be expressed as a separable differential equation for \(F(\epsilon|\Gamma)\),
\begin{equation}
    \frac{\rd F}{\rd \Gamma} e^{-F} = f(\epsilon| \Gamma) \tilde{q}_\ell(\Gamma|0).
\end{equation}
Using the initial condition \(F(T|0) = 0\), the solution is
\begin{equation}
    1-e^{-F(\epsilon|\Gamma)} =  \int_0^\Gamma \rd \gamma\, f(\epsilon| \gamma) \tilde{q}_\ell(\gamma|0),
\end{equation}
or
\begin{equation}
    F(\epsilon| \Gamma) = - \log\left[1- \int_0^\Gamma \rd \gamma\, f(\epsilon| \gamma) \tilde{q}_\ell(\gamma|0)\right].
\end{equation}
Differentiating and dividing by \(f(0| \Gamma)\) recovers \(g(\Gamma)\),
\begin{equation}\label{eqn:gsolAppendix}
    g(\Gamma) = \frac{\tilde{q}_\ell(\Gamma|0)}{1- \int_0^\Gamma \rd \gamma\, f(\epsilon| \gamma) \tilde{q}_\ell(\gamma|0)}.
\end{equation}
Recall that \(g(\Gamma) = Q(\epsilon|\Gamma) = w \rhodec(w_\Gamma)/N\), so that Eq.~\eqref{eqn:gsolAppendix} is precisely telling us the distribution of decimated elements.

To analyze this solution, we must restore the specific form of \(f\) corresponding to each of \autoref{sec:FunctionalFlowConstruction} and \autoref{sec:FunctionalFlowInstability}.
The function \(f\) is the Laplace transform of \(P\) appearing in Eq.~\eqref{eqn:qlFlowAppendix}. For the form of \(P\) in \autoref{sec:FunctionalFlowConstruction}, we can express
\begin{equation}
    f(\epsilon|\Gamma) = 2(\mathbb{E}[c^{1/\epsilon}] + \mathbb{E}[|s|^{1/\epsilon}] - 1),
\end{equation}
where \(c = \cos \tfrac{\eta}{2}\), \(s = \sin \tfrac{\eta}{2}\), and the expectation value is over a narrow range of \(\gamma \in [\Gamma, \Gamma + \rd \Gamma)\). 
In \autoref{sec:FunctionalFlowConstruction}, we take \(\epsilon \to 0\). As \(c < 1\) and \(s < 1\) whenever \(w>0\) and the energy denominator in Eq.~\eqref{eqn:RotationAngle} is finite, we have that \(c^{1/\epsilon} \to 0\) and \(s^{1/\epsilon} \to 0\), so \(f(0|\Gamma) = -2\), independent of \(\Gamma\). Substituting this gives Eq.~\eqref{eqn:gsol}, which does not demonstrate a proliferation of resonances.

In \autoref{sec:FunctionalFlowInstability}, we make a small-angle approximation, which results in 
\begin{equation}
    f(\epsilon|\Gamma) = 2 \mathbb{E}[|s|^{1/\epsilon}],
\end{equation}
and we take \(\epsilon>0\) to be small but nonzero.
In terms of the probability distribution \(P_s(s|\Gamma)\), we have
\begin{equation}
    \mathbb{E}[ |s|^{1/\epsilon} \mid \Gamma ]
    = \int_{-1/\sqrt{2}}^{1/\sqrt{2}}  |s|^{1/\epsilon} P_s(s; \Gamma) \,\rd s.
\end{equation}
The bounds on the integral come from our choice of branch for the rotation angle \(\eta\).
This integral can be estimated for small \(\epsilon\) using Laplace's method, assuming that there is a finite probability density \(0<P(1/\sqrt{2}| \Gamma) + P(-1/\sqrt{2}| \Gamma) \sim A e^{-\Gamma}\).
This is saying that there is a finite probability density of resonances, proportional to \(w\), as justified by Eq.~\eqref{eqn:ResonanceProbability}.
Then Laplace's method gives
\begin{align}
    \mathbb{E}[ |s|^{1/\epsilon} \mid \Gamma ] &\sim 2e^{-\log(2)/(2\epsilon)}[P(1/\sqrt{2}; \Gamma) + P(-1/\sqrt{2}; \Gamma)] \epsilon \nonumber\\
    &\sim A e^{-\Gamma-\log(2)/(2\epsilon)} \epsilon \nonumber\\
    &=: A_\epsilon e^{-\Gamma}/2.
\end{align}
Note that \(A_\epsilon\) is independent of \(\Gamma\), but exponentially small in \(1/\epsilon\).
Substituting this back into Eq.~\eqref{eqn:gsolAppendix} gives Eq.~\eqref{eqn:PerturbedSolutionMain}, which diverges for initial conditions with \(\theta <0\).

\section{Jacobi for a Gaussian random matrix}
\label{app:GOE}

We analyze the Jacobi algorithm when applied to a random matrix from the Gaussian orthogonal ensemble (GOE), relevant for the behavior of \(\rhodec(w)\) and \(\theta(w)\) in the dense regime.

\begin{figure}
    \centering
    \includegraphics[width=\linewidth]{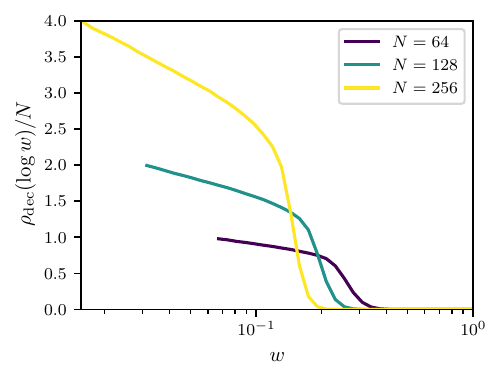}
    \caption{$\rhodec(\log w)/N$ in GOE matrices with $N=64,\, 128,\,256$. The curves are stopped at the level-spacing scale $w = 4/N$. The results are obtained by averaging over $1000$ realizations for $N=256$, $5000$ for $N=128$, and $10000$ for $N=64$.}
    \label{fig:GOE_rhodec}
\end{figure}

\begin{figure}
    \centering
    \includegraphics[width=\linewidth]{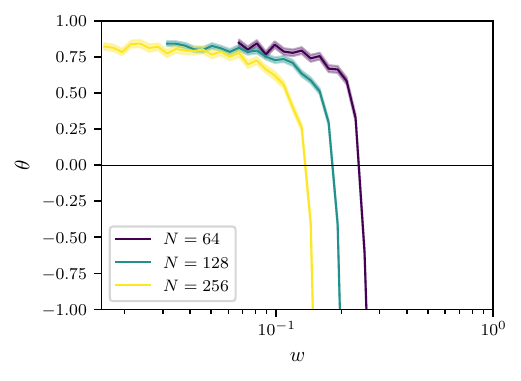}
    \caption{Flow of $\theta$ in a GOE matrix with $N=64,\, 128,\,256$, showing that the approach to $\theta = 1$ is a feature present also in random matrices and related to the approach to the level-spacing scale, occurring at $w=4/N$, where the curves end.}
    \label{fig:GOE_theta}
\end{figure}

The numerical results for \(\rhodec(\log w) = w \rhodec(w)\) are shown in \autoref{fig:GOE_rhodec}. 
We observe that the maximum value of \(w\) scales as \(1/\sqrt{N}\), as expected from the fact that the bare, computational basis matrix elements of a GOE matrix scale as \(1/\sqrt{N}\).
More interestingly, after a brief transient, we see that \(\rhodec(\log w)\) becomes only very weakly dependent on \(w\) (note the linear scale in the vertical axis of \autoref{fig:GOE_rhodec}).
This is confirmed by numerical extraction of \(\theta(w)\), as shown in \autoref{fig:GOE_theta}. 
We find that \(\theta(w) \approx 1\) for \(w < 1/\sqrt{N}\). 
This coincides with the small \(w\) behavior observed in the models of \autoref{sec:numerics}. This supports the notion that the rise of \(\theta(w)\) to roughly \(1\) is a signature of the dense regime.

Indeed, this value of \(\theta\) coincides with the known worst-case performance of the Jacobi algorithm.
Denoting the off-diagonal norm at the \(n\)th step of the Jacobi algorithm by 
\begin{equation}
    V_n^2 = \sum_{i \neq j} |H_{i_n j_n}|^2,
\end{equation}
we have that \(w_n^2 \geq V^2_n/N^2\) (the mean of the squared matrix element is smaller than the maximum).
After a single decimation, we have
\begin{equation}
    V^2_{n+1} = V_n^2 - 2 w^2 \leq V_n^2(1-2N^{-2}) \leq V_n^2 e^{-2/N^2},
\end{equation}
and thus \(V_n^2 \leq N J^2 e^{-2n/N^2}\), where \(NJ^2 = V_0^2\). This inequality is saturated (the off-diagonal norm decreases as slowly as possible) when \(w_n^2 = V_n^2/N^2\), in which case we have
\begin{multline}
    w_n = \frac{J}{\sqrt{N}} e^{-n/N^2}\\
    \implies
    \rhodec(w) = -\frac{\rd n}{\rd w} = \frac{N^2}{w} 1_{(0,J/\sqrt{N})}(w).
\end{multline}
The worst-case scaling produces the scaling with \(N\) we expect in the dense regime, and comparing the power law in \(w^{-1}\) to \(w^{-2 + \theta}\) gives \(\theta = 1\), as we observe in numerics.

\section{MBL Jacobi flow from single-particle basis}
\label{app:MBL_Anderson}

\begin{figure}
    \centering
    \includegraphics[width=\linewidth]{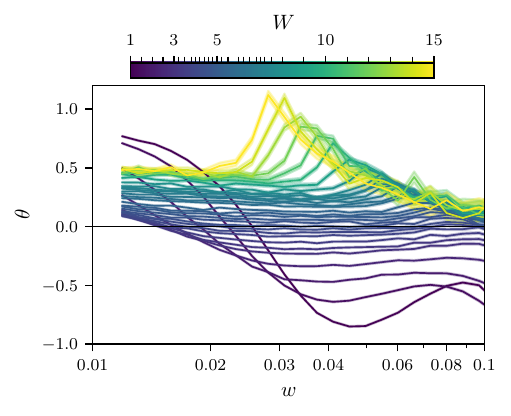}
    \includegraphics[width=\linewidth]{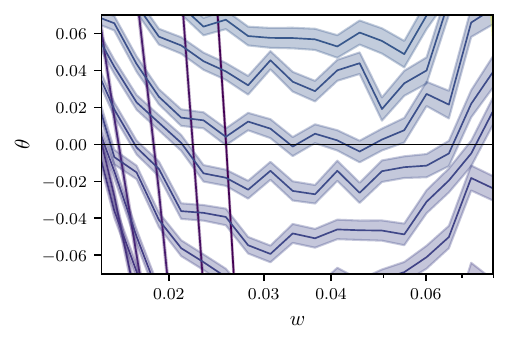}
    \caption{Jacobi flow of the resonance exponent $\theta(w)$ for the random-field XXZ chain at $\Delta=1$, with the Hamiltonian initially expressed in the many-body eigenbasis of the noninteracting $\Delta=0$ model, for $L=12$. At weak disorder, $\theta(w)$ flows toward negative values as $w$ decreases, while at strong disorder it remains positive over the accessible range. The zoomed-in view (bottom panel) shows intermediate-disorder curves that begin at positive $\theta$ and subsequently cross to negative values, signaling resonance proliferation. The early part of the flow at large $w$ depends on the starting basis: at weak disorder it resembles the GOE flow of Fig.~\ref{fig:GOE_theta}, whereas at strong disorder it is closer to the flow obtained in the $S^z$ basis.}
    \label{fig:XX_Anderson}
\end{figure}

The results for the random-field XXZ chain presented in Sec.~\ref{sec:models_XXZ} were obtained by initializing the Jacobi algorithm in the $S^z$ product-state basis. To test the dependence of the flow on this choice, we repeated the calculation for the isotropic model, $\Delta=1$, after expressing its Hamiltonian in the many-body eigenbasis of the noninteracting $\Delta=0$ model obtained from exact diagonalization.

The resulting flow of $\theta(w)$ is shown in Fig.~\ref{fig:XX_Anderson} for chains of length $L=12$. The same qualitative features discussed in the main text starting from the $S^z$ basis are present. At weak disorder, $\theta(w)$ flows toward negative values as $w$ decreases, signaling the onset of the dense, thermalizing regime. At strong disorder, the curves instead remain positive over the accessible range of $w$. The enlarged view also shows an intermediate range of disorder in which $\theta$ is initially positive but crosses zero and becomes negative later in the flow. Thus, within the accessible system sizes, the resonance-proliferation behavior reported in the main text does not depend qualitatively on whether the Jacobi algorithm is initialized in the $S^z$ basis or the $\Delta=0$ basis.

The starting basis does affect the initial part of the flow, at large $w$. At weak disorder, the single-particle localization length of the $\Delta=0$ model is large compared with the system size $L=12$. The Hamiltonian in this basis therefore has a random-matrix-like structure on the available finite system, even though the noninteracting model is localized in the thermodynamic limit. Accordingly, the initial flow resembles that found for GOE matrices in Fig.~12. At strong disorder, the Anderson orbitals are more strongly localized, and the $\Delta=0$ basis is closer to the $S^z$ product-state basis. The initial flow is then much closer to that obtained from the $S^z$ basis. These results show that the large-\(w\) transient is basis dependent, while the subsequent qualitative flow associated with resonance proliferation remains unchanged.

\bibliography{references}

\end{document}